\documentclass[a4paper,11pt]{article}
\pdfoutput=1 

\usepackage{jheppub} 

\usepackage[T1]{fontenc} 

 \usepackage{bm}
 \usepackage{amsmath}
\usepackage{graphicx}
\usepackage{subfig}
\usepackage{multirow}
\usepackage{makecell}
\usepackage{amsmath}
\usepackage{braket}
\usepackage{slashed}
\usepackage{multicol}
\usepackage{multirow}

\usepackage{soul}

\title{Measurement of CKM element $|V_{cb}|$ from $W$ boson decays at the future Higgs factories}

\author[a,b,e]{Hao Liang,}
\author[c]{LingFeng Li\protect\footnotemark[1],}
\author[a,b,d]{Yongfeng Zhu,}
\author[a,b]{Xiaoyan Shen,}
\author[a,b]{Manqi Ruan\protect\footnotemark[1],}

\footnotetext[1]{Corresponding authors.}
\affiliation[a]{Institute of High Energy Physics, Chinese Academy of Sciences, 19B Yuquan Road, Beijing 100049, China}
\affiliation[b]{University of Chinese Academy of Sciences, 19A Yuquan Road, Beijing 100049, China}
\affiliation[c]{Department of physics, Brown University, Providence, RI 02912, USA}
\affiliation[d]{State Key Laboratory of Nuclear Physics and Technology, School of Physics, Peking University, Beijing 100871, China}
\affiliation[e]{Vanderbilt University Institute of Imaging Science, Vanderbilt University Medical Center, Nashville, TN 37232, USA}

\emailAdd{hao.liang@vumc.org}
\emailAdd{lingfeng\_li@brown.edu}
\emailAdd{zhuyongfeng@pku.edu.cn}
\emailAdd{shenxy@ihep.ac.cn}
\emailAdd{ruanmq@ihep.ac.cn}

\abstract{

This study investigates the precision measurement of the CKM matrix element $|V_{cb}|$ through semileptonic $WW$ events at future Higgs factories, i.e., FCC-ee, ILC, C$^3$, and CEPC. We use full detector simulation to generate the $WW \to \ell \nu cb$ signal events and various backgrounds at $\sqrt{s} = 240$ GeV with unpolarized beams. The relative statistical uncertainties of $|V_{cb}|$ are projected to be 0.91\% for the muon channel and 1.2\% for the electron channel, assuming a baseline integrated luminosity of 5 ab$^{-1}$. The sensitivities at other Higgs factory scenarios are also projected. Possible contributors to systematic uncertainties are discussed, with the most prominent one being the systematics of flavor-tagging and mistagging rates. Combining with $WW$ threshold runs, the relative systematic uncertainty can be further reduced.

}


\begin{document}
\maketitle


\section{Introduction}

The measurement of $|V_{cb}|$, the magnitude of the Cabibbo-Kobayashi-Maskawa (CKM) matrix element~\cite{Cabibbo:1963yz,Kobayashi:1973fv} representing the transition between $c$ and $b$ quarks, is essential for understanding the weak interactions in the Standard Model (SM) and probing potential new physics. The $|V_{cb}|$ value is fundamental for charged current decays that dominate the total width of $b$-hadrons and control the flavor-changing neutral current (FCNC) processes. The latter are sensitive to new physics contributions due to their suppressed amplitudes in the SM. 
A persistent discrepancy between $|V_{cb}|$ values from inclusive and exclusive b-hadron decays remains a longstanding puzzle in flavor physics.~\cite{Bigi:2017jbd,Bernlochner:2017xyx,Bernlochner:2019ldg,Ricciardi:2019zph,Iguro:2020cpg,Gambino:2020jvv,Ricciardi:2021shl}. Currently, the exclusive $b$-hadron decays yield an average $|V_{cb}|$ of $(3.910\pm 0.050)\times 10^{-2}$, whereas inclusive $B$ decays result in $(4.219\pm 0.078)\times 10^{-2}$~\cite{HFLAV:2022pwe}.  The potential violation of CKM unitarity 
may indicate new physics interacting with the SM flavor sector~\cite{Czarnecki:2004cw,Belfatto:2019swo,Charles:2020dfl,Branco:2021vhs}. However, this $3\sigma$ tension is not yet conclusive evidence of new physics due to systematic uncertainties (both perturbative and non-perturbative) inherent in hadronic observables~\cite{Leibovich:1997tu,Leibovich:1997em,Ligeti:2014kia,MILC:2015uhg,Bernlochner:2016bci,Bernlochner:2017jxt}.

Measuring the decay width in $W\to cb$ transitions provides an alternative approach to determine $|V_{cb}|$ at an energy scale $\gg \Lambda_{\rm QCD}$, minimizing non-perturbative corrections. Furthermore, by extracting $|V_{cb}|$ values from the ratio of $\Gamma(W\to cb)/\Gamma(W\to qq)$, many theoretical and experimental uncertainties will cancel out, such as the leading perturbative QCD corrections. Such a measurement also provides independent consistency checks since its systematic uncertainties differ from those in $b$-hadron decays. Evaluating the CKM elements at electroweak energy scales also enhances the experiment's sensitivity to potential new physics introduced at a UV scale much higher than the electroweak scale~\cite{Descotes-Genon:2018foz}.

Despite the clarity of the physics picture of obtaining $|V_{cb}|$ from $W$ boson, no experimental result of $|V_{cb}|$ measured through $W$ boson decay is available. The only CKM element measurements via W boson decay are the $|V_{cs}|$ measurements at LEP~\cite{DELPHI:1998hlc,ALEPH:1999djo,OPAL:2000jsa}, while the $|V_{cb}|$ measurement at LEP is impractical due to the small $\text{Br}(W\to cb)$ value ($6\times 10^{-4}$ in the SM) and the limited statistics of $W$. Recent studies have explored $|V_{cb}|$ measurement through on-shell $W$ bosons at the LHC~\cite{Harrison:2018bqi,Choi:2021esu,Vladimirov:2022pdd}. However, the projected relative sensitivity of $|V_{cb}|$ at the LHC is only of $ \mathcal{O}(10\%)$, which is too weak to resolve the current tension. Even in the upcoming high-luminosity phase, it is unlikely that this method will achieve a relative statistical uncertainty below $2\%$. Meanwhile, systematic error from jet tagging and large QCD backgrounds will be challenging to be mitigated at the same level (see e.g.,~\cite{ATLAS:2018xcf,ATLAS:2024tnr}).

The recent proposition of employing future $e^+e^-$ colliders offers unique opportunities for the measurement of $|V_{cb}|$ through $W$ decays with high statistics and low background. These next-generation colliders are also referred to as Higgs factories, including the $ee$ phase of Future Circular Collider (FCC-$ee$)~\cite{FCC:2018evy,Bernardi:2022hny,deBlas:2024bmz}, the International Linear Collider (ILC)~\cite{Behnke:2013xla,ILCInternationalDevelopmentTeam:2022izu}, the Cool Copper Collider (C$^3$)~\cite{Bai:2021rdg}, and the Circular Electron Positron Collider (CEPC)~\cite{CEPCStudyGroup:2018ghi,CEPCPhysicsStudyGroup:2022uwl}. The flavor physics potential at these future lepton colliders are recently evaluated in many phenomenological studies~\cite{Marzocca:2024dsz,Zheng:2020ult,Li:2022tov,Li:2022tlo,Aleksan:2021gii,Aleksan:2021fbx,Amhis:2021cfy,Kamenik:2017ghi,Li:2020bvr,Monteil:2021ith,Chrzaszcz:2021nuk,Dam:2018rfz,Qin:2017aju,Li:2018cod,Calibbi:2021pyh,Aleksan:2024jwm,Altmannshofer:2023tsa,Amhis:2023mpj,Ho:2022ipo,Zuo:2023dzn}, though most of them focus on the lower energy runs. During their Higgs factory or $WW$ threshold operations, these facilities will provide a sample of $\gtrsim\mathcal{O}(10^8)$ $W$ decays and the unique opportunity to measure $|V_{cb}|$ through hadronic $W$ decays.\footnote{The $|V_{cb}^{W\to cb}|$ obtained this way is defined at the scale of $m_W$, which may take different values from $|V_{cb}|$ observed from hadron physics, especially in the presence of new physics~\cite{Descotes-Genon:2018foz}. However, within SM, $|V_{cb}|$ can be treated as a constant below the weak scale~\cite{ParticleDataGroup:2022pth}, and we take the short-hand notation of $|V_{cb}|$ without ambiguity.} This work focuses on the measurement at the unpolarized Higgs factory mode with $\sqrt{s}=240-250$~GeV, where the $e^+e^-\to W^+W^-$ production rate is significant with a high integrated luminosity. The sensitivities from other experimental conditions will be projected from these baseline results.

The paper is structured as follows. Sec.~\ref{sec:detsim} introduces the CEPC baseline detector model and the simulation software. Sec.~\ref{sec:evtsel} describes the event selection and flavor tagging performance. Sec.~\ref{sec:res} introduces the method to obtain the $|V_{cb}|$ and project for various experimental scenarios. In Sec.~\ref{sec:system}, various sources of systematic effects will be estimated. We then briefly summarize in Sec.~\ref{sec:conc}.

\section{Detector Simulation and Data Samples}
\label{sec:detsim}

Measuring of $|V_{cb}|$ targeting a sub-percent precision is challenging as the experiment need to collect a large enough clean signal samples while controlling the systematic effects to be no bigger than the target precision. In this study, we adopt full detector simulations to provide conservative but realistic projections for future studies. For detector simulation, \textsc{MokkaPlus}~\cite{MoradeFreitas:2002kj}, a \textsc{GEANT 4}~\cite{GEANT4:2002zbu}-based simulation framework is used, with the baseline detector profile derived from the International Large Detector (ILD)~\cite{CEPCStudyGroup:2018ghi,Einhaus:2023npl}. 
 In particular, the track reconstruction is based on \textsc{Clupatra}~\cite{Gaede:2014aza}, and the particle flow reconstruction is based on the \textsc{Arbor}~\cite{Ruan:2013rkk, Ruan:2018yrh} algorithm. \textsc{Marlin}~\cite{Gaede:2006pj} and \textsc{LCIO}~\cite{Gaede:2003ip} from ilcsoft are used for data management.
We use $ee$-kt algorithm for the jet clustering via the toolkit of \textsc{LCFIPlus}~\cite{Suehara:2015ura}, with all isolated leptons (see Sec.~\ref{ssec:muon} for details) removed from the clustering process.
For jet flavor tagging, we use the methodology of jet origin identification~\cite{Bedeschi:2022rnj,Liang:2023yyi}, applying \textsc{ParticleNet} algorithm~\cite{Qu:2019gqs} on the physics event reconstructed with \textsc{Arbor} to distinguish jets into 10 different categories corresponding to 10 different classes of (anti-)quarks.

To generate signal and background samples, we follow the approach in~\cite{An:2018dwb} of simulation, focusing on the unpolarized beam case with $P\equiv (P_{e^-},P_{e^+})=(0,0)$ with $\sqrt{s}=240$~GeV. All background and signal samples in this work are generated by \textsc{Whizard}~\cite{Kilian:2007gr} and \textsc{Pythia 6}~\cite{Sjostrand:2006za}. To avoid large hadronic backgrounds, we focus on the semileptonic $WW\to cb \ell\nu$ mode as the signal. In this case, an isolated lepton with significant energy can distinguish our signal from most non-$WW$ backgrounds. Most backgrounds come from hadronic $W$ decays other than $W\to cb$ with very similar kinematics. They are, hence, the ``irreducible" background for our study and can only be distinguished from the signals via jet flavor tagging. On the contrary, the non-$WW$ backgrounds can be efficiently removed with proper kinematic selection rules. Our simulated backgrounds fall into the following categories: \textbf{Semileptonic $WW$ decays:} they are almost identical to signal events but with different $W$ hadronic decays. After the flavor-tagging cuts, they will be the dominant background. \textbf{Other $WW$ events:} this category includes other $WW$ decays, including $WW\to \tau\nu qq$ events with various $\tau$ decays, semileptonic $WW$ decays with the wrong lepton flavor, and fully hadronic $WW$ decays. \textbf{Other four fermion events:} the category summarizes four fermion processes that are not $WW$ or Higgs decays. Examples include various $ZZ$ decays, $e^+e^-\to e^+e^- Z$ single production, and the interference between $WW$ and $ZZ$ four-fermin final states. 
\textbf{Higgs processes:} it includes $\ell\ell+H$, $\nu\nu+H$, and $qq+H$ events simulated. \textbf{Two fermion events:} the class stands for Drell-Yan production of fermion pairs with or without strong initial state radiation photons, and excludes neutrino production processes.



\section{Event Selection and Analysis}

\label{sec:evtsel}
\subsection{Kinematic-based Selections for the Muon Channel}
\label{ssec:muon}

\begin{table}[htbp!]
    \centering
    \setlength{\tabcolsep}{1.3pt}
{\footnotesize

    \begin{tabular}{|c|ccc|ccc|cccc|ccccccccc}
    \hline
              & $\mu \nu cb$ &  $\mu \nu cd/s$ &   $\mu \nu qq_{\rm{other}}$ &                $\mu_\tau \nu cb$ & $\mu_\tau \nu cd/s$ & $\mu_\tau \nu qq_{\rm other}$ & $WW_{\rm{other}}$ & 4$f_{\rm{other}}$ & Higgs & 2$f$\\
    \hline
    
No selections &40.3{{\tiny K}}&24.2{{\tiny M}}&24.2{{\tiny M}}&7.73{\tiny K}&4.2{\tiny M}&4.2{\tiny M}&194.3{{\tiny M}}&133.2{{\tiny M}}&4.07{\tiny M}&1.78G\\
$E_{{\rm L}\ell} > 12~{\rm GeV}$ &37.9{{\tiny K}}&22.6{{\tiny M}}&22.6{{\tiny M}}&5.59{\tiny K}&2.98{\tiny M}&2.97{\tiny M}&20.8{{\tiny M}}&17.5{{\tiny M}}&645{{\tiny K}}&152.0{{\tiny M}}\\
$R_{{\rm L}\ell} > 0.85$ &35.6{{\tiny K}}&21.3{{\tiny M}}&21.3{{\tiny M}}&5.04{\tiny K}&2.75{\tiny M}&2.75{\tiny M}&16.5{{\tiny M}}&10.0{{\tiny M}}&311{{\tiny K}}&105.2{{\tiny M}}\\
$q_{{\rm L}\ell}\cos(\theta_{{\rm L}\ell}) < 0$ &31.8{{\tiny K}}&19.0{{\tiny M}}&19.0{{\tiny M}}&4.57{\tiny K}&2.49{\tiny M}&2.49{\tiny M}&14.4{{\tiny M}}&7.0{\tiny M}&157{{\tiny K}}&67.8{{\tiny M}}\\
2nd isolation $\ell$ veto &31.8{{\tiny K}}&19.0{{\tiny M}}&19.0{{\tiny M}}&4.57{\tiny K}&2.49{\tiny M}&2.49{\tiny M}&13.8{{\tiny M}}&3.16{\tiny M}&117{{\tiny K}}&24.2{{\tiny M}}\\
Multiplicity $ \ge 15$ &31.8{{\tiny K}}&18.9{{\tiny M}}&18.8{{\tiny M}}&4.57{\tiny K}&2.48{\tiny M}&2.47{\tiny M}&104{{\tiny K}}&623{{\tiny K}}&100{{\tiny K}}&131{{\tiny K}}\\
$\slashed{p}_T > 9.5~{\rm GeV}$ &30.5{{\tiny K}}&18.1{{\tiny M}}&18.1{{\tiny M}}&4.26{\tiny K}&2.32{\tiny M}&2.32{\tiny M}&90.7{{\tiny K}}&234{{\tiny K}}&78.1{{\tiny K}}&71.5{{\tiny K}}\\
$M_{\rm jets}>65~{\rm GeV}$ &28.5{{\tiny K}}&17.6{{\tiny M}}&17.7{{\tiny M}}&4.04{\tiny K}&2.25{\tiny M}&2.27{\tiny M}&67.7{{\tiny K}}&223{{\tiny K}}&71.9{{\tiny K}}&19.9{{\tiny K}}\\
$M_{\rm jets}<88~{\rm GeV}$ &23.6{{\tiny K}}&14.0{{\tiny M}}&13.8{{\tiny M}}&3.41{\tiny K}&1.82{\tiny M}&1.8{\tiny M}&42.2{{\tiny K}}&16.7{{\tiny K}}&5.63{\tiny K}&3.99{\tiny K}\\
$M_{\rm jets, recoil}<115~{\rm GeV}$ &19.8{{\tiny K}}&12.8{{\tiny M}}&12.9{{\tiny M}}&2.9{\tiny K}&1.68{\tiny M}&1.69{\tiny M}&38.4{{\tiny K}}&9.97{\tiny K}&477&3.43{\tiny K}\\
$M_{{\rm L}\ell{\rm S}\ell} < 75~{\rm GeV} / c^2$ &19.3{{\tiny K}}&12.7{{\tiny M}}&12.8{{\tiny M}}&2.89{\tiny K}&1.68{\tiny M}&1.69{\tiny M}&38.4{{\tiny K}}&8.96{\tiny K}&361&3.43{\tiny K}\\
$\epsilon_{\rm kin} (\%)$ &47.8&52.6&52.8&37.4&39.9&40.1&0.019&0.006&0.008&0.0002\\

\hline
    \end{tabular}
    }

    \caption{Event cut flow for the muon channel at the unpolarized Higgs factory, with all yields normalized to 20 ab$^{-1}$. See the text for detailed definitions of each cut. 
After event selection, the main components of $4f_\text{other}$ are from $ZZ$.    
}
\label{tab:evtselct1}
\end{table}

\begin{figure}[htbp]
    \centering
    \includegraphics[width=7 cm]{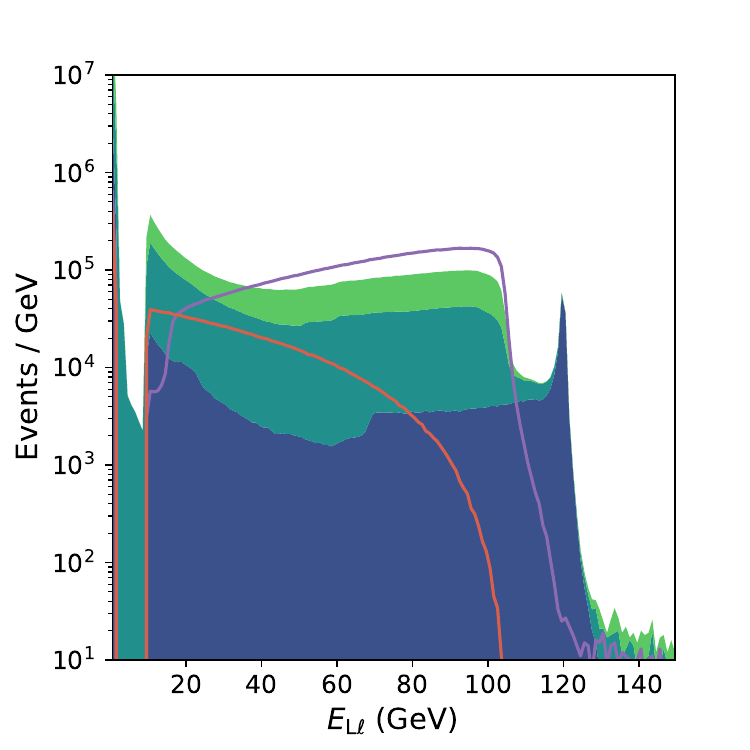}
    \includegraphics[width=7 cm]{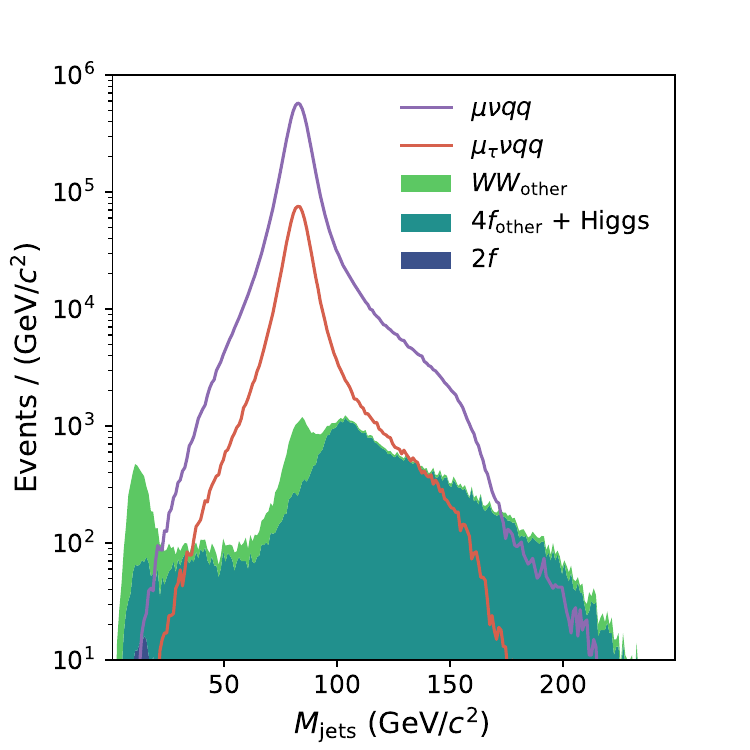}
    \caption{The distribution of the leading lepton energy $E_{\rm{L}\ell}$ and jet invariant mass $M_{\rm jets}$ in cut chain. Since the kinematic cuts do not discriminate different jet flavor effectively, all semileptonic leptonic $WW$ events are plotted together. All samples shown for each distribution are the ones right before the relevant step in the cut flow.
The 10 GeV cutoff in the left panel is due to suboptimal particle
identification (PID) for low-energy leptons in the current version of construction
tools. However, it does not affect the isolated leptons with energies greater
than 10 GeV, which is the region of interest.
}
    \label{fig:cutvariables}
\end{figure}

We first focus on the muon channel with only one hard isolated muon in the final state. 
To select final state of semileptonic $WW$ decays with a muon from background events, the following cuts are applied to MC samples.
During the process, the $WW\rightarrow \tau(\to \mu\nu\bar{\nu})cb$ events are also recognized as the signal to maximize the signal yield. 
In the rest of this work, we will take the short hand notation of $\mu_\tau$ or $e_\tau$ for leptons from leptonic $\tau$ decays.

\begin{enumerate}
    \item Leading (in energy) muon shall be an isolated muon. We identify events with isolated lepton as: (1) the leading muon have an energy ($E_{\rm{L} \ell}$ ) greater than 12 GeV. (2) the energy fraction of the leading muon itself within a cone of $30^ {\circ}$ around it ($R_{\rm{L} \ell}$) is  greater than 85\% .

    \item Leptons generated from $WW$ decay experience strong forward-backward asymmetry, with their direction and charge correlated with the beam. We therefore require the product of the leading lepton's charge ($q_{\rm{L}\ell}$) and the cosine of its polar angle ($\theta_{L\ell}$) to be negative.
    
    \item Sub-leading muon shall not be another isolated muon of the opposite sign. Therefore, the sub-leading lepton shall satisfy either of the following criteria: its energy ($E_{\rm{S}\ell}$) smaller than 22 GeV, or its energy fraction of muon in the cone around it within an angle of $30^ {\circ}$ ($E_{\rm{S}\ell}$) less than 0.9, or it 
    is an electron, or it has the same charge as the leading muon.
    
    \item Missing $p_T$ ($\slashed{p}_T$). We select events with missing $p_T$ greater than 9.5~GeV.

    \item The multiplicity of reconstructed particles with energy greater than 0.5~GeV.
    We select events with multiplicities $\ge$ 15 to veto the backgrounds with fully leptonic final states. 
    
    \item Invariant masses. To select the $W$ from the low missing energy backgrounds like $ZZ$ and $ZH$, we select events with the invariant mass of the dijet in the range from 65 to 88 GeV. The invariant mass of the leading and subleading muon (if any), must be less than 75 GeV.

    \item Invariant masses from recoil. To further exclude the background without hard neutrinos,
    the recoil mass of jets $M_{\rm jets, recoil}\equiv \sqrt{(p_{ee}^\mu-\sum p_{\rm jets}^\mu)^2}$ is required to be less than 115~GeV, where $p_{ee}$ is the 4-momentum of the $e^+e^-$ initial state.
    
\end{enumerate}

The distributions of the key cut variables are plotted in Fig.~\ref{fig:cutvariables}, in which the cuts before the corresponding variable have been applied. Since the cuts are insensitive to quark flavors in the final state, we do not specify jet flavor in these plots.\footnote{In practice, heavy quark decays give rise to neutrinos and extra leptons, thus altering relevant cut efficiencies. However, from our simulation, the differences in terms of overall efficiency are less than 1\% for non-$b$ events, while for events with $b$ quarks, the difference could be as large as 10\%.} The other distributions are plotted in App.~\ref{app:kinematic}. Note that all cuts are based on kinematic features with no explicit jet flavor arguments. For signal events, the average efficiency of such kinematic cuts ($\epsilon_{\rm kin}$) is about 47\% considering both $\mu$ and $\mu_\tau$ signals. The detailed cut flows of each signal and background mode are listed in Table.~\ref{tab:evtselct1}. Since the most significant difference appears in the $WW_{\rm othters}$ category, we provide a more detailed comparison in App.~\ref{app:eventsstatistics}.

\subsection{Kinematic-based Selections for the Electron Channel}

\begin{table}[htbp!]
    \centering
    \setlength{\tabcolsep}{1.3pt}
{\footnotesize

    \begin{tabular}{|c|ccc|ccc|cccc|}
    \hline
              & $e \nu cb$ &  $e \nu cd/s$ &   $e \nu qq_{\rm{other}}$ &                $e_\tau \nu cb$ & $e_\tau \nu cd/s$ & $e_\tau \nu qq_{\rm other}$ & $WW_{\rm{other}}$ & 4$f_{\rm{other}}$ & Higgs & 2$f$\\
    \hline
    
No selections &47.1{{\tiny K}}&26.1{{\tiny M}}&26.1{{\tiny M}}&7.83{\tiny K}&4.32{\tiny M}&4.33{\tiny M}&190.3{{\tiny M}}&133.2{{\tiny M}}&4.07{\tiny M}&1.78G\\
$E_{{\rm L}\ell} > 12~{\rm GeV}$ &44.4{{\tiny K}}&24.6{{\tiny M}}&24.8{{\tiny M}}&6.19{\tiny K}&3.44{\tiny M}&3.54{\tiny M}&115.8{{\tiny M}}&84.3{{\tiny M}}&743{{\tiny K}}&924.4{{\tiny M}}\\
$R_{{\rm L}\ell} > 0.85$ &38.7{{\tiny K}}&21.4{{\tiny M}}&21.2{{\tiny M}}&4.16{\tiny K}&2.28{\tiny M}&2.24{\tiny M}&21.5{{\tiny M}}&23.8{{\tiny M}}&259{{\tiny K}}&478.8{{\tiny M}}\\
$q_{{\rm L}\ell}\cos(\theta_{{\rm L}\ell}) < 0$ &34.6{{\tiny K}}&19.1{{\tiny M}}&19.0{{\tiny M}}&3.83{\tiny K}&2.08{\tiny M}&2.04{\tiny M}&18.7{{\tiny M}}&20.2{{\tiny M}}&133{{\tiny K}}&455.1{{\tiny M}}\\
2nd isolation $\ell$ veto &34.6{{\tiny K}}&19.1{{\tiny M}}&19.0{{\tiny M}}&3.83{\tiny K}&2.07{\tiny M}&2.03{\tiny M}&17.9{{\tiny M}}&12.6{{\tiny M}}&100{{\tiny K}}&166.5{{\tiny M}}\\
Multiplicity $ \ge 15$ &34.6{{\tiny K}}&19.1{{\tiny M}}&18.9{{\tiny M}}&3.83{\tiny K}&2.07{\tiny M}&2.02{\tiny M}&3.73{\tiny M}&2.59{\tiny M}&80.0{{\tiny K}}&4.32{\tiny M}\\
$\slashed{p}_T > 9.5~{\rm GeV}$ &33.0{{\tiny K}}&18.1{{\tiny M}}&18.0{{\tiny M}}&3.55{\tiny K}&1.93{\tiny M}&1.89{\tiny M}&3.27{\tiny M}&786{{\tiny K}}&55.5{{\tiny K}}&1.24{\tiny M}\\
$M_{\rm jets}>65~{\rm GeV}$ &30.7{{\tiny K}}&17.5{{\tiny M}}&17.6{{\tiny M}}&3.29{\tiny K}&1.86{\tiny M}&1.84{\tiny M}&3.14{\tiny M}&611{{\tiny K}}&48.9{{\tiny K}}&878{{\tiny K}}\\
$M_{\rm jets}<88~{\rm GeV}$ &24.9{{\tiny K}}&13.8{{\tiny M}}&13.5{{\tiny M}}&2.73{\tiny K}&1.49{\tiny M}&1.45{\tiny M}&1.66{\tiny M}&55.8{{\tiny K}}&4.5{\tiny K}&214{{\tiny K}}\\
$M_{\rm jets, recoil}<115~{\rm GeV}$ &19.9{{\tiny K}}&12.1{{\tiny M}}&12.0{{\tiny M}}&2.3{\tiny K}&1.37{\tiny M}&1.35{\tiny M}&1.5{\tiny M}&32.9{{\tiny K}}&350&187{{\tiny K}}\\
$M_{{\rm L}\ell{\rm S}\ell} < 75~{\rm GeV}$ &18.2{{\tiny K}}&10.5{{\tiny M}}&10.2{{\tiny M}}&2.23{\tiny K}&1.33{\tiny M}&1.3{\tiny M}&1.46{\tiny M}&26.5{{\tiny K}}&276&115{{\tiny K}}\\
$\epsilon_{\rm kin} (\%)$ &38.6&40.1&39.0&28.5&30.7&30.0&0.76&0.019&0.006&0.006\\

\hline
    \end{tabular}
    }

    \caption{Similar to Table~\ref{tab:evtselct1} but for the electron channel. 
    }
\label{tab:evtselct2}
\end{table}

\begin{figure}[h!]
    \centering
    \includegraphics[scale=0.5]{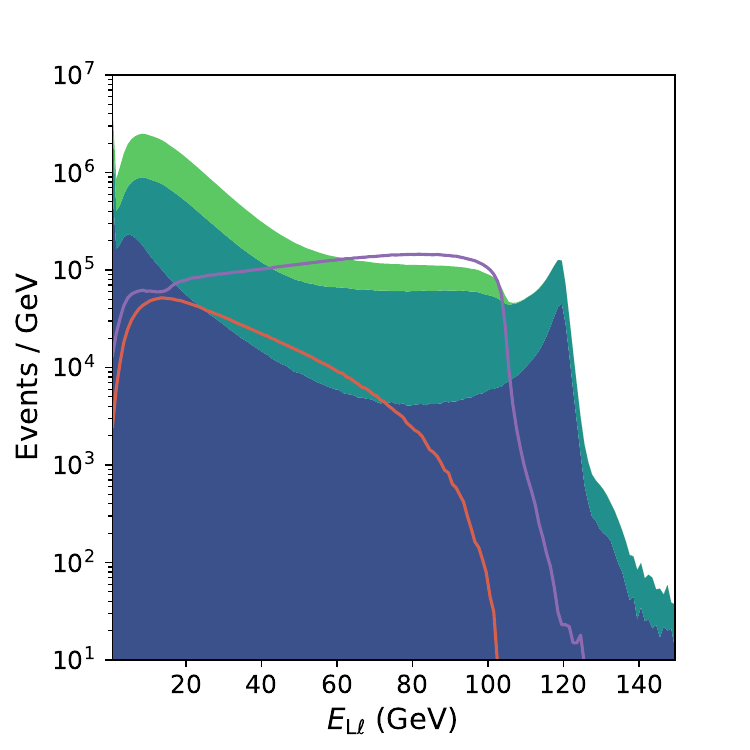}
    \includegraphics[scale=0.5]{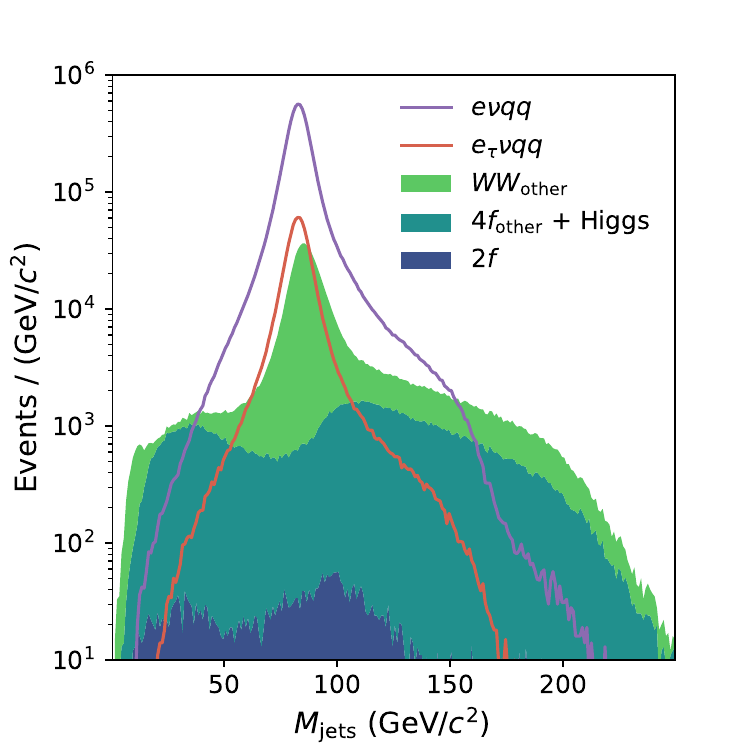}
    \caption{Distributions of kinematic variables in the cut flow fo the electron channel, similar to Fig.~\ref{fig:cutvariables}.}
    \label{fig:cutvariables_ele}
\end{figure}

For the electron final states from $WW$ semileptonic decays, we apply a similar cut flow to simulated samples in Sec.~\ref{ssec:muon} but replacing muons with electrons. 
The cut flow is listed in Table~\ref{tab:evtselct2}, behaving similar to the one in Table~\ref{tab:evtselct1}.\footnote{Since the simulation work also includes $e^+e^-\to e\nu W$ events with the $e\nu$ not from on-shell $W$ decays, the production rates are slightly larger than the counterparts in Sec.~\ref{ssec:muon}. We will not specify the difference in the rest of this paper since it also helps the $|V_{cb}|$ measurement.} This channel faces challenging electron particle identification and the background events with electrons recoiled from the beam, such as $e^+e^- \to e^+e^- Z$ events.
In particular, we find there is a significant chance of 4\% that the hadronic decay of tau leptons is misidentified as isolated electrons in $\tau\nu_\tau q\bar{q}$ samples with current lepton PID.
Their impacts are visible in Fig.~\ref{fig:cutvariables_ele}. Both the signal efficiency and signal-to-noise ratio are thus moderately downgraded compared to the muon channel, making it a subleading contributor to $|V_{cb}|$ measurement.  

\subsection{Flavor Tagging}
The measurement of $V_{cb}$ from hadronic $W$ decays is only possible if one can discern $b$ and $c$ jets from light-quark-initiated ones. Therefore, the flavor tagging and corresponding mistagging rates are crucial for the measurement's precision and accuracy. As the benchmark study, we estimate the flavor tagging performance with a method similar to the ones in~\cite{Bedeschi:2022rnj,Liang:2023yyi}. The individual jet tagging efficiency is based on the ParticleNet algorithm~\cite{Qu:2019gqs} calibrated from $Z\to q\bar{q}$ events at the $Z$-pole, where the neuron network classifies a jet into ten possible categories, i.e., $b,~c,~s,~u,~d$ and correspond antiquarks. Each jet from an event is then associated with a 10-dimensional score correlated with another jet. This method's benchmark $b$-tagging efficiency is $\sim 91\%$, with a $\sim 4\%$ mistagging rate from $c$ and $\sim 0.5\%$ from lighter quarks.
A more conservative flavor tagging method, directly using the full simulation inputs, is also performed for cross-check; see App.~\ref{app:traditionaltagging} for more details.
\footnote{Compared to the benchmark reported in the FCC-$ee$ study~\cite{Bedeschi:2022rnj}, which is also based on the ParticleNet algorithm~\cite{Qu:2019gqs}, the algorithm we adopt has a few differences. Firstly, the training and validation sets are based on full detector simulation instead of fast simulation. Also, the algorithm's output covers the sign of quarks in the hard process beside their flavor. Both can lead to a more conservative flavor tagging performance.}

\begin{figure}
    \centering
    \includegraphics[height=4.8cm]{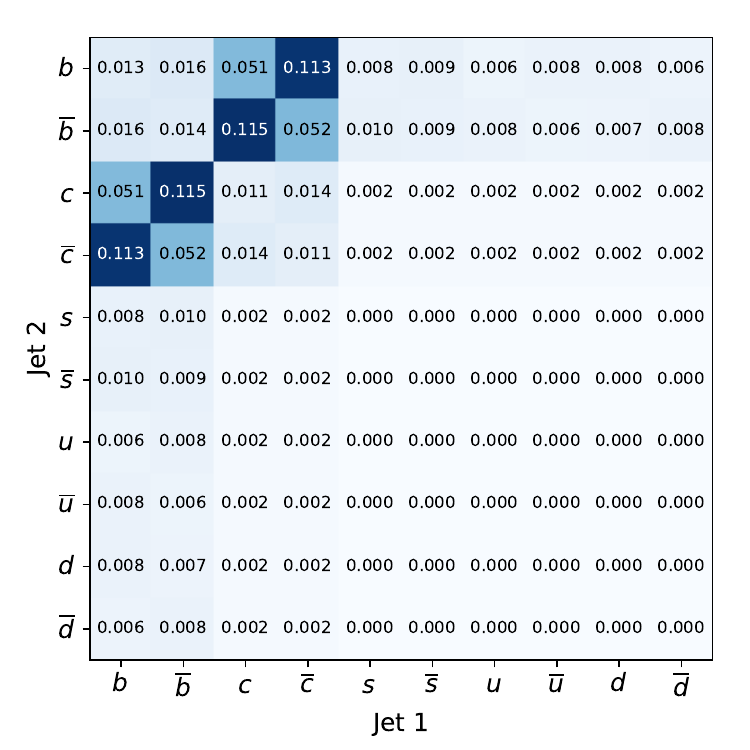}
    \includegraphics[height=4.8cm]{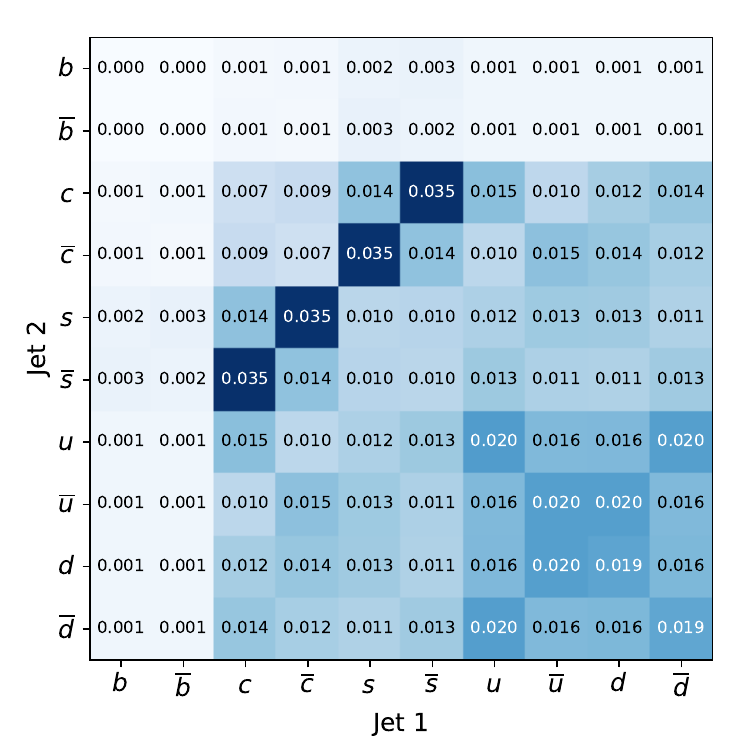}
     \includegraphics[height=4.8cm]{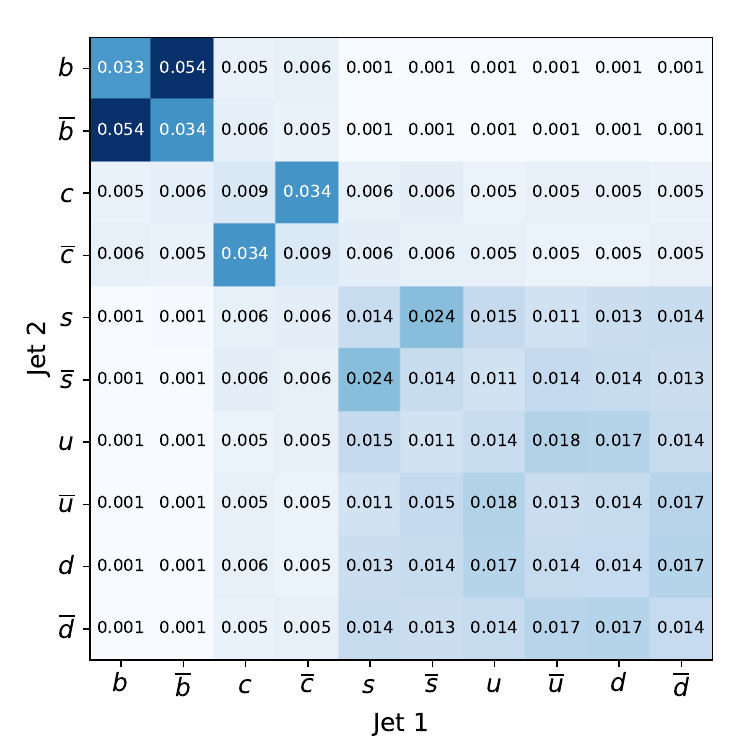}
    \caption{The average tensor products of the two jets' 10 dimensional scores for $WW\to \mu\nu cb$ (left), other $WW$ background events (middle), and the rest of backgrounds (right).}
    \label{fig:10times10matrix}
\end{figure}

In Fig.~\ref{fig:10times10matrix}, we plot the average tensor products of the two jets' 10-dimensional scores. The $W\to cb$ signal events plotted in the left panel have high scores in $c\bar{b}$ or $b\bar{c}$ combinations, followed by $cb$ and $\bar{c}\bar{b}$ due to charge mis-ID. The other $W$ decay backgrounds in the middle panel have clear features in $c\bar{s}$ or $s\bar{c}$ combinations, while the $u\bar{d}$ and $d\bar{u}$ feature is blurred as it is challenging to tag these light jets. It is also clear that the major contributors are same-flavor pairs like $b\bar{b}$ or $c\bar{c}$ from the distribution of remaining backgrounds in the right panel. The jet tagging performance of $e\nu qq$ final states is similar since the algorithm does not take isolated lepton information as inputs. The main difference is the varying background composition. While the flavor structure in $WW$ processes are governed by $W$ decay branching ratios, it differs moderately between non-$WW$ background components. Numerically, we find that the relative differences between the two lepton channels in the entries of Fig.~\ref{fig:10times10matrix} are less than 10\%.

\section{Extraction of Signal Strengths and Effective Statistics}
\label{sec:res}
\subsection{Projected $V_{cb}$ Sensitivities}

\begin{figure}[h!]
    \centering
    \includegraphics[height=6.5 cm]{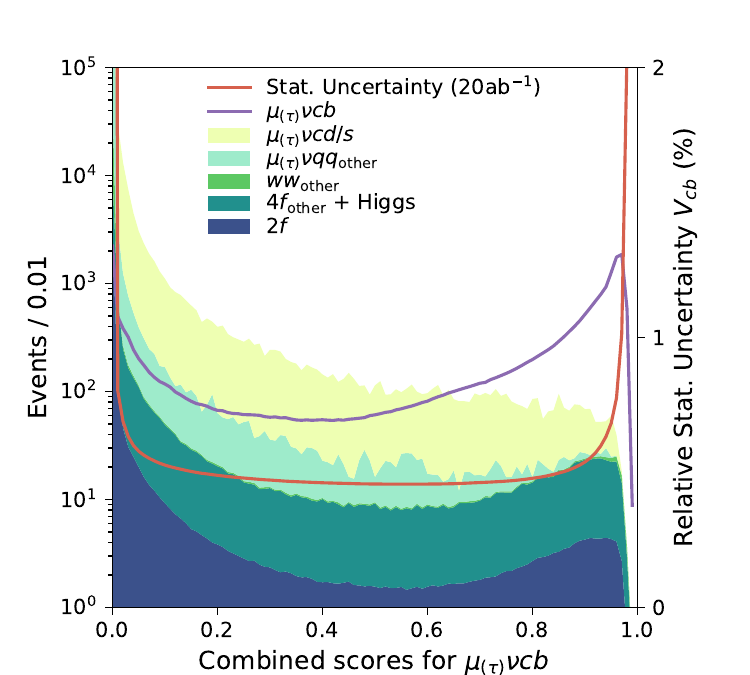}
    \includegraphics[height=6.5 cm]{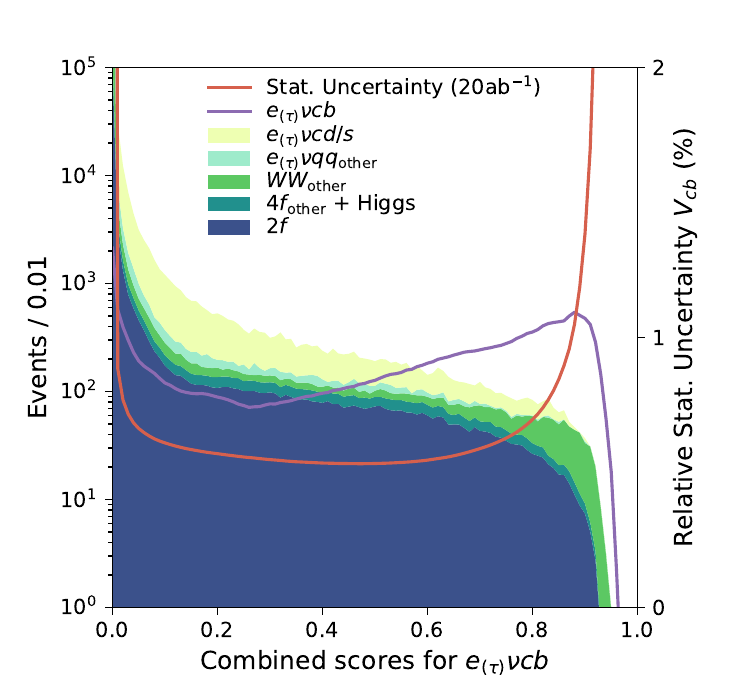}
    \caption{The BDT score distribution of signal and backgrounds in: the muon channel (left) and electron channel (right). The red curve indicates the projected statistical relative sensitivity estimated from Eq.~\ref{eq:stat} assuming a luminosity of 20~ab$^{-1}$.}
    \label{fig:BDToutput}
\end{figure}

We introduce a boost decision tree- (BDT)-based classifier to identify $W\to cb$ signals from the flavor tagging scores shown in Fig.~\ref{fig:10times10matrix}. The classifier takes the 10-dimensional tagging scores from each jet as its input (20 in total) to identify signals from various background events selected by kinematic cuts. Its outputs for both $WW\to \mu\nu qq$ and $WW\to e\nu qq$ are plotted in Fig.~\ref{fig:BDToutput}, in which the effectiveness of the flavor tagging algorithm is apparent.

At the leading order, the above measurement and $||V_{cb}|$ value has the relation $R_{cb}\equiv \frac{\text{BR}(W\to cb)}{\sum \text{BR}(W\to qq)}\simeq \frac{|V_{cb}|^2}{2}$. The statistical uncertainties for the baseline unpolarized Higgs factory mode are obtained via the relation:
\begin{footnotesize}
\begin{equation}
    \label{eq:stat}
    \frac{\delta |V_{cb}|}{|V_{cb}|}\bigg|_{\rm stat}= \frac{1}{2}\frac{\delta R_{cb}}{R_{cb}}\bigg|_{\rm stat}= \frac{1}{2}\bigg[\frac{\delta\text{BR}(WW\to \ell \nu cb)}{\text{BR}(WW\to \ell \nu cb)}\bigoplus \frac{\delta\text{BR}(WW\to \ell \nu qq)}{\text{BR}(WW\to \ell \nu qq)}\bigg]\bigg|_{\rm stat}\simeq \frac{1}{2\sqrt{\epsilon \rho N_{WW\to \ell \nu cb} }}~,
\end{equation}
\end{footnotesize}
where $N_{WW\to\ell \nu cb}$ is the raw event number produced at the Higgs factory, $\epsilon$ is the overall signal efficiency to the final signal region, and $\rho$ is the purity defined as the signal yield divided by all events in the signal region. The second term in the squared bracket induced by the overall $WW$ semileptonic decay rate is always subdominant and thus neglected safely. The product of the form $\epsilon \rho N$ determines the statistical uncertainty and works as the ``effective statistics" of the channel. The corresponding results are shown in the red curves in Fig.~\ref{fig:BDToutput} for an unpolarized Higgs factory with 20~ab$^{-1}$. The optimal results are obtained by choosing output thresholds that ensure high purity without sacrificing signal efficiency substantially. 
We thus define the signal region according to those optimal cuts, with the statistics listed in Table~\ref{tab:statistics_abs}. 

The signal efficiency $\epsilon_{\rm tag}$  and purity $\rho$ for both signal modes are summarized in Table \ref{tab:statistics}. Thanks to the high efficiency of kinematic and flavor-tagging cuts, the backgrounds are suppressed in both channels at the cost of signal efficiencies. Here, the muon channel enjoys a higher purity, with about $25\%$ backgrounds stemming from non-$WW$ events. To reduce systematic uncertainties, we choose the BDT working point that delivers high purity by sacrificing some tagging efficiency for the electron channel. The relative contribution from the non-$WW$ backgrounds increases to $\sim 40\%$ due to the large $e^+e^- \to e^+e^-+2f$ inclusive backgrounds. Summarizing all the above, the raw and effective statistics of both signal channels for the unpolarized mode per ab$^{-1}$ are shown in Table.~\ref{tab:statistics}. The sensitivity in various Higgs factory scenarios will be projected based on these numbers. {{The efficiency for each event class are further detailed in App.~\ref{app:eventsstatistics}}}

\begin{table}[htb!]
    \centering
    \begin{footnotesize}
    \begin{tabular}{|cccccc|}
    \hline
      Mode  & \makecell[c]{$N_{WW\to \ell \nu cb}$}/ab$^{-1}$ &  $\epsilon_{\rm kin}$ & $\epsilon_{\rm tag}$ & $\rho$ & \makecell[c]{Eff. stat.}/ab$^{-1}$ \\
      \hline
       $\mu_{(\tau)}\nu cb$, $\sqrt{s}=$240~GeV, $P=(0,0)$ &  2410 &   0.462 & 0.642  & 0.854 & 6.1 $\times 10^2$ \\
       $e_{(\tau)}\nu cb$, $\sqrt{s}=$ 240~GeV, $P=(0,0)$ &  2746 & 0.372 & 0.439 & 0.824  &    3.7 $\times 10^2$\\
    \hline
    \end{tabular}
    \end{footnotesize} 
    \caption{Projected effective statistics and statistical uncertainties for each benchmark $WW\to \ell\nu qq$ mode, including $\tau\to \ell\nu\nu$ contributions. The signal yield $N_{WW\to \ell \nu cb}$ are normalized to a luminosity of 1~ab$^{-1}$}
\label{tab:statistics}  
\end{table}

\subsection{Different Scenarios of Higgs Factory}
\textbf{$W$ factory mode} Many proposed Many proposed Higgs factory facilities will operate with the $W$ factory mode at $\sqrt{s} \simeq 2 m_W$~\cite{Behnke:2013xla,FCC:2018evy,CEPCStudyGroup:2018ghi} to measure $W$ boson properties precisely. Though ideal for studying properties like $m_W$~\cite{Wilson:2016hne,Azzurri:2021yvl}, here we argue that it may not be the dominant scenario for $|V_{cb}|$ measurement due to the limited $WW$ production rate near the $2m_W$ threshold, varying between 1-10~pb, which is lower than the rate $\sim 17$~pb at $\sqrt{s}=240$~GeV. Moreover, $W$ hadronic decays significantly contribute to the background, which can only be removed by the tagging algorithm. Even removing all non $WW$ backgrounds will not improve the effective statistics of $|V_{cb}|$ measurements for the muon channels. However, the lower non-$WW$ background level can reduce the impact from systematic effects, as discussed in~\cite{Marzocca:2024dsz} via fully hadronic final states with fast simulation. For this semileptonic analysis, we tentatively rescale the non-$WW$ backgrounds by a uniform factor of 0.5 to incorporate the effect. This approach is in rough accordince with the LEP study~\cite{OPAL:2007ytu}, where the signal-to-background ratio of $WW\to qq\ell \nu$ measurement increases by$\sim 30\%$ when $\sqrt{s}$ changes from 161.3 to 206.6~GeV.

A baseline of $10^8~W$ bosons (or $5\times 10^7~WW$ events) near the $WW$ threshold are chosen, corresponding to an integrated luminosity of $\sim 10$ ab$^{-1}$ with a typical $\sigma_{e^+e^-\to W^+W^-} \simeq 5-6$~pb near the $2m_W$ threshold. The projected sensitivities on $|V_{cb}|$ by both lepton modes are shown in Table~\ref{tab:scenarios}, including the combined ones with the unpolarized Higgs factory runs. In addition, the fully hadronic $WW$ decays will also contribute to the $|V_{cb}|$ measurement given the lower non-$WW$ background level, as demonstrated in the recent study~\cite{Marzocca:2024dsz}. We hope the potential and systematics of this method can be fully evaluated with dedicated simulation.



\textbf{Polarized Higgs factory mode} To project the results for linear Higgs factories with typical beam polarization $P=(-0.8,0.3)$ with $\sqrt{s}=250$~GeV, we assume the signal and background efficiencies are approximately the same as the unpolarized version. Also, a minor shift of $\sqrt{s}$ would not drastically affect efficiencies and cross sections. For instance, the $WW$ cross section only reduces by $2\%$ when $\sqrt{s}$ increases by 10~GeV. From the Whizard simulation, the signal cross section increases by a factor of $\sim 2.3$, very close to the theoretical maximum of (1+0.8)(1+0.3)=2.34 since the $e^-_Le^+_R$ contribution dominates the event rate. Other SM backgrounds, such as the $ZZ$ or $2f$ rates, increase by $\lesssim 2$ times, while the general 4-fermion background still grows by a factor of 2.3~\cite{Yan:2016xyx,Karl:2019hes}. Given that most backgrounds for measuring $|V_{cb}|$ are irreducible, it is then reasonable to take the effective $\epsilon \times \rho$ to be the same as the corresponding values in Table~\ref{tab:statistics} to be conservative. In Table.~\ref{tab:scenarios}, we summarize the projections for various Higgs factory scenarios from different signal modes. 

Note that with a similar approach, one can also obtain the statistical uncertainties of the $R_{cs}\equiv \frac{\text{BR}(W\to cs)}{\sum \text{BR}(W\to qq)}$ to measure the $|V_{cs}|$ value, which is currently known at a relative precision of $\sim 0.6\%$~\cite{HFLAV:2022pwe}. The estimated statistical uncertainty of $|V_{cs}|$ at most Higgs factory scenarios is well below 0.1\%. In this case, the systematic uncertainties will dominate.

\begin{table}[htb!]
    \centering
    \begin{footnotesize}
    \begin{tabular}{|cccc|}
    \hline
      Mode  & $WW\to\mu\nu cb$ & $WW\to e\nu cb$  & Combined \\
      \hline
       Unpolarized, Baseline (5~ab$^{-1}$) &  0.91\% &  1.2\%   &  0.72\% \\
       Unpolarized, Extended (20~ab$^{-1}$) &  0.45\% & 0.58\%  & 0.36\% \\
       $WW$ Threshold ($5\times 10^7~WW$) &  1.2\% & 1.6\% & 0.95\% \\
       Unpolarized, Baseline + $WW$ &  0.72\% & 0.96\% & 0.57\%  \\
      Unpolarized, Extended + $WW$ &  0.42\% & 0.56\% &   0.33\% \\ 
        Polarized, Baseline (0.5~ab$^{-1}$) &  1.9\% & 2.4\% &  1.5\% \\
        Polarized, Extended (2~ab$^{-1}$) &  0.94\% & 1.2\% &  0.75\% \\

    \hline
    \end{tabular}
    \caption{The $|V_{cb}|$ relative sensitivity projected for each Higgs factory scenarios. For unpolarized Higgs factories operating at $\sqrt{s}=240$~GeV, the baseline and extended scenarios correspond to 5 and 20 ab$^{-1}$, respectively. The $WW$ threshold run corresponds to $5\times 10^7$ $ee\to WW$ events. We also present the sensitivities when the two modes are combined. For polarized Higgs factories, we take the polarization benchmark $P = (-0.8,0.3)$ and $\sqrt{s} = 250$~GeV. The baseline and extended scenarios correspond to integrated luminosity of 0.5 and 2 ab$^{-1}$, respectively. }
    \label{tab:scenarios}
    \end{footnotesize}
\end{table}

\section{Systematic Effects}
\label{sec:system}
As a precision test of the SM, the low statistic uncertainties of $|V_{cb}|$ measurements are demonstrated in many different Higgs factory scenarios above. For such a precision measurement, the systematic effects must be carefully evaluated, both theoretical and experimental. Theoretical uncertainties include errors when translating the $R_{cb}$ measurement to a $|V_{cb}|$ value, such as perturbative corrections or uncertainties of other input parameters. Instead, experimental uncertainties trace possible differences from the $R_{cb}$ signal strength measured from its truth value induced by MC simulation, detector effects, and analysis algorithms. We will estimate the contribution from various terms and discuss the necessary conditions to have them under control.

\subsection{Theoretical Uncertainties}

In Sec.~\ref{sec:res}, the steps of $R_{cb}\equiv \frac{\Gamma(W\to cb)}{\sum \Gamma(W\to qq)}$ measurement is developed with details. Though $|V_{cb}|^2=R_{cb}\sum_{f=d,s,b}(|V_{uf}|^2+|V_{cf}|^2)$ is a good approximation at the leading order, perturbative corrections on $W$ decay widths must be included in practice. Generically, they originate from two sources: (1) uncertainties of parameters used for the calculation, such as quark masses or other CKM matrix elements; (2) intrinsic uncertainties introduced by higher order corrections unaccounted. Beyond the leading order relation, higher order corrections of $R_{cb}$ are only relevant when they break the $Z_3$ symmetry between generations, e.g., the quark masses. In the following, we will try to set a reasonable upper limit for each term from known results.

\textbf{Parametric uncertainties from quark masses} The $m_f$ contributions to the theoretical uncertainties start at the leading order but are suppressed by the hierarchy of $m_{c,b}^2/m_W^2<0.3\%$. The values of $m_{c,b}$, though scheme dependant, are known at 1\% level~\cite{ParticleDataGroup:2022pth}. Considering future progressions, they thus will introduce a relative parametric uncertainty $\leq 1\times 10^{-4}$.


\textbf{Parametric uncertainties from top mass}  The large $m_t$ significantly breaks the $Z_3$ symmetry, but its effect starts to appear at EW one-loop corrections. From~\cite{Kniehl:2000rb}, the relative correction shall be $\mathcal{O}\bigg(\frac{\alpha |V_{tb}|^2}{\pi \sin^2\theta_W}\times\frac{\delta m_t}{m_t}\bigg)$. Meanwhile, the top mass uncertainty $\delta_t$ is already as small as $\sim 0.5$~GeV from the LHC measurements~\cite{ParticleDataGroup:2022pth}. The value will further improve during the HL-LHC era and the high energy runs of Higgs factories. The corresponding parametric uncertainty of $|V_{cb}|$ reduces to $\leq 1\times 10^{-4}$, which is compatible with its effect on $Z$ decays~\cite{Dubovyk:2018rlg,Dubovyk:2019szj}.

\textbf{Parametric uncertainties from CKM matrix elements} Notice the leading-term relation between $|V_{cb}|^2=\sum_{f=d,s,b}(|V_{uf}|^2+|V_{cf}|^2)R_{cb}$, impacts from CKM matrix experimental uncertainties vanishes with CKM unitarity. The CKM uncertainty reappears in higher-order corrections, however, further suppressed by the $Z_3$ symmetry-breaking hierarchies of $\text{Max}\bigg(\frac{m_b^2}{m_W^2},\frac{\alpha}{\pi \sin^2\theta_W}\bigg)\lesssim 0.01$. Even assuming the current uncertainty of CKM matrix elements other than $|V_{cb}|$, the parametric uncertainty induced will be at most $1\times 10^{-4}$~\cite{dEnterria:2020cpv} assuming CKM unitarity. 

\textbf{Remaining parametric uncertainties from QCD and EW theories} In general, those input parameters like $m_Z$, $m_t$, $\sin \theta_W$, or $\alpha(m_Z)$ are constrained well below $10^{-2}$~\cite{ParticleDataGroup:2022pth} level alrady. Combined with the $Z_3$ symmetry suppression of $\mathcal{O}(10^{-2})$, their effect will be subdominant. The largest contribution is expected from the relative uncertainty of $\alpha_s$, currently known at $\sim 3\%$. It will be, again, narrowed down by about an order of magnitude from the $Z$ pole run with $10^{12}$ $Z$ at future lepton colliders~\cite{dEnterria:2020cpv}. Taking the $Z$ decay results~\cite{Dubovyk:2018rlg,Dubovyk:2019szj} as reference, we expect the $|V_{cb}|$ parametric uncertainty of this type to be no larger than $1\times 10^{-4}$.

\textbf{Intrinsic theoretical uncertainties from higher order corrections} The latest calculation of hadronic $W$ decays~\cite{dEnterria:2020cpv} including the quark mass effects up to $\mathcal{O}(\alpha_s)$~\cite{Almasy:2008ep}, and $\mathcal{O}(\alpha\alpha_s)$~\cite{Kara:2013dua}. The intrinsic systematic uncertainty will emerge at higher orders. Currently, the most prominent terms are the $\mathcal{O}(\alpha^2)$ ones, which could bring a relative uncertainty as big as $1\times 10^{-3}$. since similar calculations exist for Z decays, it is reasonable to expect $\mathcal{O}(\alpha^2)$ calculations within the next decade, given the need for precision W physics programs. In addition, given that the quadratic quark mass terms are available at $\mathcal{O}(\alpha_s^2)$ and $\mathcal{O}(\alpha_s^3)$~\cite{Chetyrkin:1996hm} already, the $Z$ decay results~\cite{Dubovyk:2019szj} suggest the intrinsic uncertainty of $|V_{cb}|$ to be $\leq 3\times 10^{-4}$, coming from $\mathcal{O}(\alpha^2\alpha_s)$ and $\mathcal{O}(\alpha\alpha_s^2)$ corrections. The number may further reduce to $\leq 1\times 10^{-4}$ if the quark mass effects in both terms above are addressed properly. 



Summarizing the above, the overall theoretical uncertainty of $|V_{cb}|$ determination from $R_{cb}$ is therefore unlikely to exceed $4\times 10^{-4}$ assuming CKM unitarity. Note that the experimental approach provides the $R_{cb}$ value as an observable, which can be reinterpreted as other quantities addressing different prospects and with even smaller uncertainties such as $|V_{cb}|^2/(\sum_i |V_{ci}|^2)$. The systematic uncertainty of $R_{cb}$ itself, however, is largely experimental as explained in the following. 

    

\subsection{Experimental Uncertainties}
\label{sec:extra}

\textbf{Flavor Tagging} 
Our study's most prominent systematic uncertainties are anticipated to arise from tagging efficiencies, i.e, performance differences between calibrated MC simulations and the experimental data. The tagging procedure affects the $R_{cb}$ measurement accuracy by introducing signal yield and background estimation uncertainties. The signal efficiency $\epsilon_{\rm tag}$ from flavor tagging, as shown in Fig.~\ref{fig:10times10matrix}, can be estimated by the product of $b$ and $c$ jet tagging efficiency. The systematic uncertainty of $\epsilon_{\rm tag}$ directly affects the $R_{cb}$ measurement. We take the averaged tagging efficiency difference between samples generated by \textsc{Herwig 7}~\cite{Bellm:2015jjp} and \textsc{Pythia 6}~\cite{Sjostrand:2006za} as a baseline estimator of flavor-tagging uncertainties. The extrapolated deviation of $b$ and $c$ tagging efficiency from MC simulation is about $0.8\%$, which applies to both the muon and electron channels. 

The systematic uncertainties introduced by backgrounds are suppressed by the high signal-to-noise ratio. However, we must stress that the algorithm's relative systematic uncertainties in terms of mistagging rates shall also be higher. We first focus on the $s\to b$ mistagging since it is the major mistag pattern for the dominant $W\to cs$ background, as shown in Fig.~\ref{fig:10times10matrix}. By applying the same method above, the relative systematic uncertainty in the $s\to b$ mistagging is estimated to be $\sim 4.5\%$. Similarly, the non-$WW$ backgrounds mostly contribute to the signal region from $b$ and $c$ pairs with $b/c$ mistagging. The estimated relative systematic uncertainty of flavor-mistagging between $b$- and $c$- jets is $\sim 3.5\%$. As rough estimators, the above numbers are based on the averaged efficiencies without explicitly including subleading systematics like the kinematic dependence or multi-jet correlations. Since the jet distributions must be precisely calibrated at future lepton colliders for many studies, we do not expect their contribution to inflate our estimation significantly.

Summarizing the above, the relative $|V_{cb}|$ systematic uncertainty from flavor tagging is $\sim 2\%$, surpassing the statistical ones in Table.~\ref{tab:scenarios} in general. However, the differences could be significantly mitigated using the abundant $e^+e^- \to q\bar{q}$ dijet data and advanced calibration algorithms. Moreover, the flavor correlation between jets may further enhance the algorithm's accuracy (see~\cite{Gambhir:2024tgs,Gouskos:2024ijk}). In the case of full calibration, the $b$ and $c$ tagging efficiencies may have their systematic uncertainty level further reduced by five times~\cite{Gouskos:2024ijk}. If the mistag uncertainties drop by the same factor, the systematic level from flavor tagging will be reduced to $\lesssim 0.3\%$ for both channels.

\textbf{Non-$WW$ background estimation} Systematic uncertainties due to background estimations are suppressed by the high signal purity achieved. A proper selection of control regions beside the signal region can further suppress the effect. 
Provided the typical background rate uncertainty for semileptonic $WW$ channels $\lesssim 10\%$ at LEP~\cite{DELPHI:1998hlc,ALEPH:1999djo,OPAL:2000jsa,OPAL:2007ytu}, the above benchmark of shall be an achievable target at future Higgs factories. 

\textbf{Kinematic selection efficiencies for semileptonic $W$ decays} From the definition of $R_{cb}$, the systematic effects from kinematic selection rules largely cancel, similar to the case of theoretical uncertainties. According to the efficiencies in Table~\ref{tab:evtselct1}, the relative efficiencies of $W\to cb$ and other hadronic $W$ decay modes differ by up to $10\%$. It is thus reasonable to assume that the systematic uncertainty of the relative efficiency is also further suppressed by an order of magnitude. The relevant systematic uncertainty estimated this way will be $\lesssim 0.4\%$, even assuming no improvement from the LEP results~\cite{OPAL:2007ytu}. In all, this term will be subdominant compared to other systematic effects and can be safely ignored.

\begin{figure}
\centering
\includegraphics[width=13 cm]{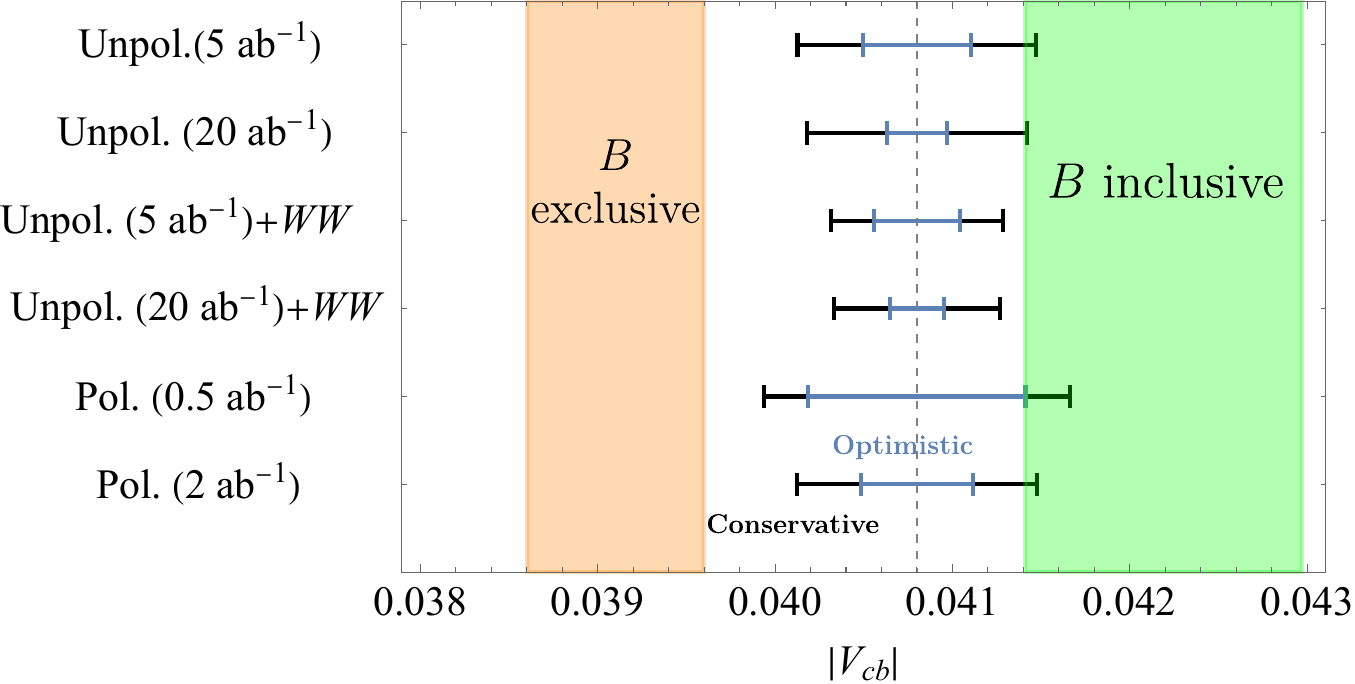}
\caption{Summary plot of $|V_{cb}|$ projected sensitivities, including systematic effects. Six major Higgs factory benchmarks from Table.~\ref{tab:syst} are presented, with the two scenarios shown as black and blue error bars. For comparison, the current $|V_{cb}|$ ranges from inclusive and exclusive $B$ decays~\cite{HFLAV:2022pwe} are also presented, while the PDG average~\cite{particle2022review} as the nominal central value for future projections.}
\label{fig:money}
\end{figure}

In summary, the experimental uncertainties of $|V_{cb}|$ measurement, together with the overall systematic effects after combining both signal channels, are shown in Table~\ref{tab:syst} and Fig.~\ref{fig:money}. In particular, we consider two different scenarios for systematic effects. The first one is conservative, taking the flavor tagging uncertainty by comparing different MC simulations and the non-$WW$ background rate uncertainty to be 2\%, which is a mild improvement compared to the LEP results~\cite{OPAL:2007ytu}. In the secondary scenario, we consider more optimistic control on tagging efficiency and background yields. In particular, the uncertainty in tagging efficiency is reduced by eight times compared to the first scenario. The background uncertainty is reduced by four times to $0.5\%$, which improves by about an order of magnitude compared to the LEP results~\cite{OPAL:2007ytu,ALEPH:2013dgf,L3:2005fft,ALEPH:2006cdc,DELPHI:2008avl}. The overall systematic uncertainty is obtained by comparing the difference between the final and statistical-only sensitivity, assuming the systematics in each measurement are uncorrelated.
It is worth mentioning that combining the measurements at the Higgs factory mode and the $WW$ threshold runs can further reduce the experimental uncertainties. The $|V_{cs}|$ measurement will enjoy an even higher signal purity. Therefore, we expect the dominant systematic contribution to arise from the uncertainty $c$ and $s$ jet tagging efficiencies with sizes comparable to those in Table.~\ref{tab:syst}.

\begin{table}[htb!]
    \centering
    \begin{footnotesize}
    \begin{tabular}{|cccc|}
    \hline
      Uncertainty  & Stat.  & Syst., Secnario 1  & Syst., Secnario 2  \\
      \hline
       Unpolarized, Baseline (5~ab$^{-1}$) &  0.72\% &  1.5\%   &  0.20\% \\
       Unpolarized, Extended (20~ab$^{-1}$) &  0.36\% &  1.5\% & 0.20\% \\
       $WW$ Threshold ($5\times 10^7~WW$) & 0.95\% & 1.5\% & 0.20\% \\
       Unpolarized, Baseline + $WW$ &  0.57\% & 1.1\% & 0.15\%  \\
      Unpolarized, Extended + $WW$ &  0.33\% & 1.1\% &   0.18\% \\ 
        Polarized, Baseline (0.5~ab$^{-1}$) &  1.5\% & 1.5\% &  0.20\% \\
        Polarized, Extended (2~ab$^{-1}$) &  0.75\% & 1.5\% &  0.20\% \\

    \hline
    \end{tabular}
    \caption{The projected $|V_{cb}|$ relative sensitivities by combining muon and electron channels. The statistical sensitivities are identical to the last column of Table.~\ref{tab:scenarios}. Systematic uncertainties are derived for two scenarios: a more conservative one (scenario 1) and an optimistic one (scenario 2). These systematic impacts come from the performance differences between the final and statistical-only sensitivity; see the text for details.}
    \label{tab:syst}
    \end{footnotesize}
\end{table}

\section{Summary}
\label{sec:conc}
Measuring CKM matrix elements is central to flavor physics and serves as an efficient probe for new physics. Measuring $|V_{cb}|$ from $W\to cb$ decays provides a complimentary way free from non-perturbative uncertainties and with independent systematic effects. Such analyses will be most effective at future Higgs factories, where high statistics and low systematic uncertainties can be achieved simultaneously.

In this work, we use full detector simulation to project the uncertainties for $|V_{cb}|$ at unpolarized Higgs factories via the observable $R_{cb} \equiv \frac{\text{BR}(W \to cb)}{\sum \text{BR}(W \to qq)}$. We focus on semileptonic $WW$ final states with a $\mu$ or $e$ with significant missing energy to ensure high signal efficiency, purity, and clear theoretical interpretation. Kinematic selection rules are proposed to discriminate $WW$ events from other SM backgrounds, including multiple four- and two-fermion events by at least four orders of magnitude with signal efficiencies $\gtrsim 40\%$. Flavor tagging cuts based on the \textsc{ParticleNet} algorithm further separate $W \to cb$ events from other hadronic $W$ decays with high purity and efficiency. The analysis projects a statistical uncertainty of $\sim 0.4-1.5\%$ for $|V_{cb}|$ when combining both leptonic channels, depending on operational plans. These results demonstrate the potential of $|V_{cb}|$ extraction at Higgs factories, comparable to $WW$ threshold runs. 

Theoretical and experimental sources of systematic uncertainties are discussed. Higher-order corrections to the measurement are suppressed in the $R_{cb}$ ratio, introducing a relative $|V_{cb}|$ uncertainty $\lesssim 0.1\%$ from current perturbative calculations. Experimental systematic effects are more significant and may dominate over statistical ones in a conservative scenario. We recognize the largest term is the uncertainty in flavor tagging due to the tagging efficiencies of $b/c$ jets and mistagging rates from lighter jets to $b/c$ jets. This is crucial as backgrounds like $WW \to \ell \nu cs/d$ can only be efficiently removed through flavor tagging. The effects from tagging and mistagging are comparable in our baseline simulations. Other systematic uncertainties from signal efficiency and background estimation will also contribute moderately. Overall, systematic uncertainties may overtake statistical ones in the worst case but could be mitigated with the help of data calibration and specific algorithms. Here, our goal is to roughly pin down the possible range of systematic effects of $|V_{cb}|$ extraction. On the other hand, a thorough study of the precise magnitude of systematics and identifying subleading contributors will be crucial and widely benefit precision physics programs at future lepton colliders. Eventually, the accurate determination $|V_{cb}|$ will only be possible if the systematics are properly modeled, as a community effort.

The measurement of $R_{cb}$ from semileptonic $WW$ measurements could be generalized for other measurements like $R_{cs}$ or $R_{cd}$, which may help further extend our understanding of precision $W$ physics. We hope the method and conclusions of this work will become useful inputs fitting well into the broader picture of joint electroweak-flavor physics program at future Higgs factories.

\section*{Acknowledgement}
We thank Wolfgang Altmannshofer, Ayres Freitas, Loukas Gouskos, Zoltan Ligeti, Manuel Szewc, and Jure Zupan for useful discussions. We also thank Yuexin Wang for contributing in the early stage of this work. HL, YZ, XS, and MR are supported by the Innovative Scientific Program of the Institute of High Energy Physics. LL is supported by DOE grant DE-SC0010010.



\appendix

\section{Kinetic Distributions in the Cut Flow}
\label{app:kinematic}
\begin{figure}[htbp]
    \centering
    \includegraphics[width=7cm]{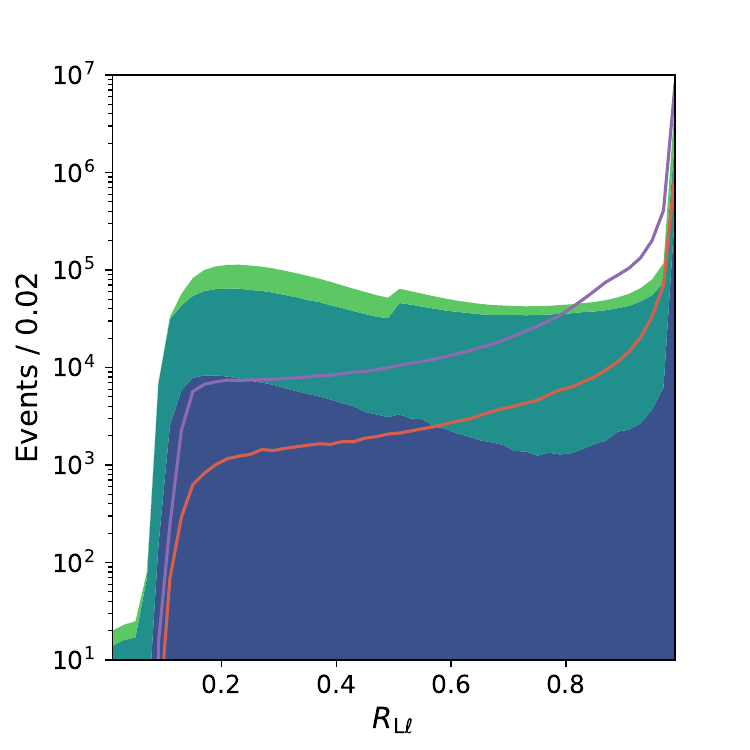}
    \includegraphics[width=7cm]{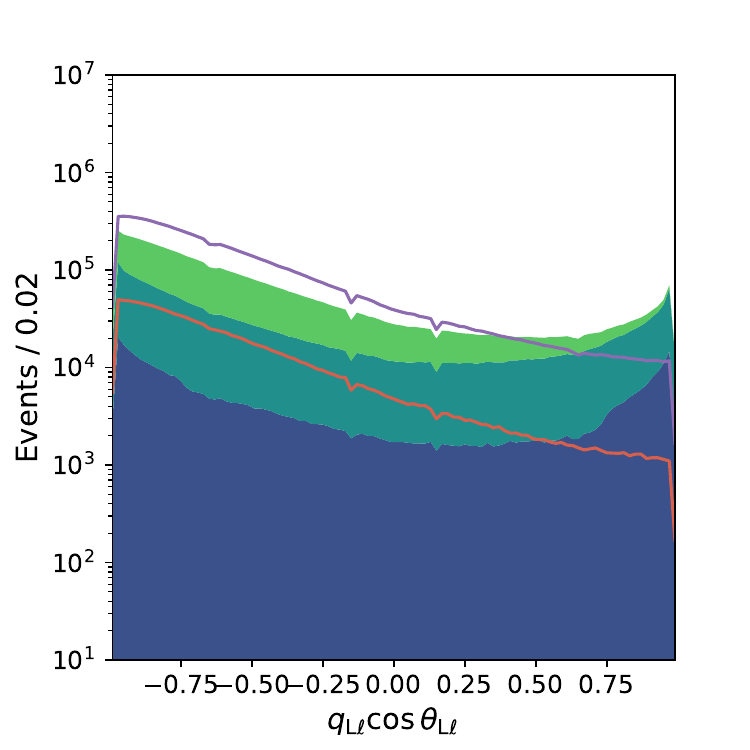}    
    \includegraphics[width=7cm]{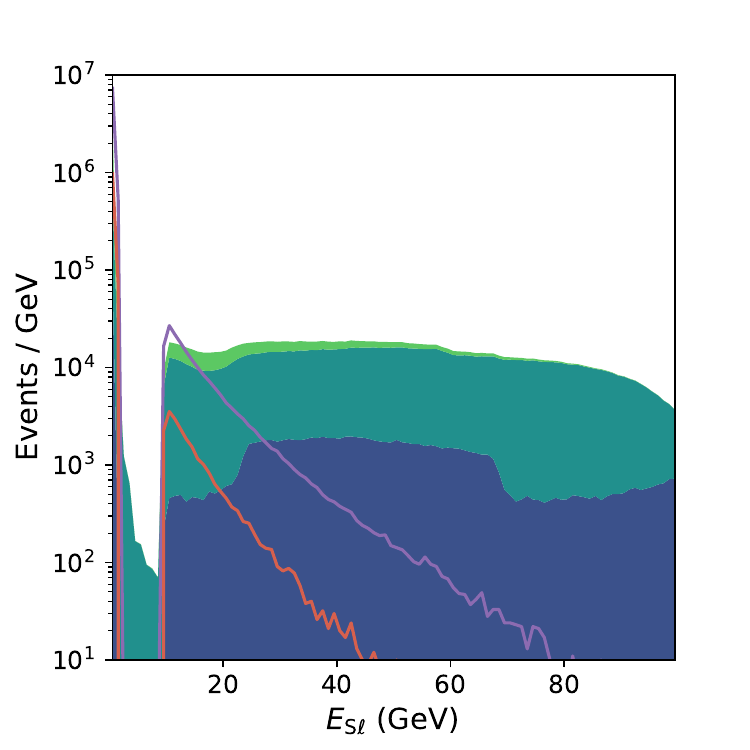}
    \includegraphics[width=7cm]{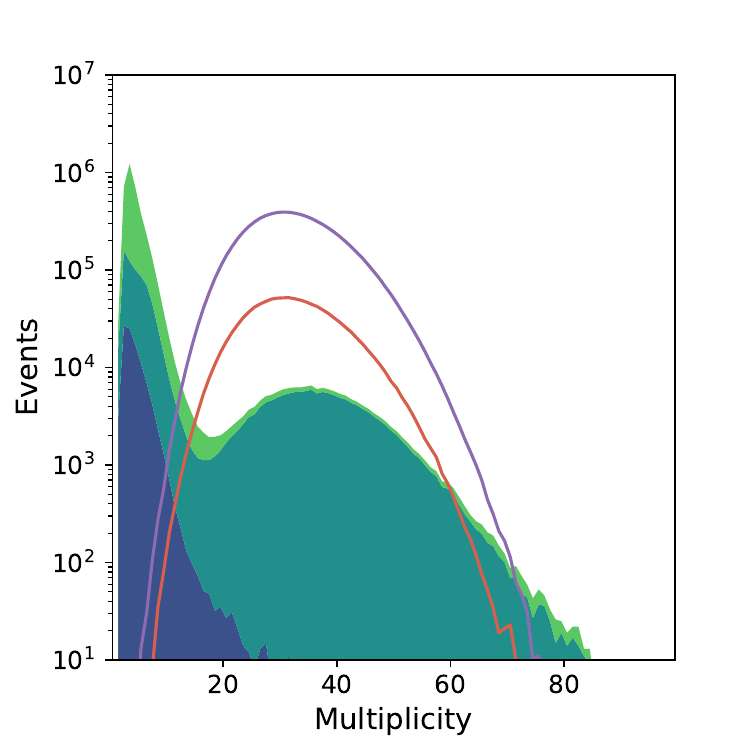}
    \includegraphics[width=7cm]{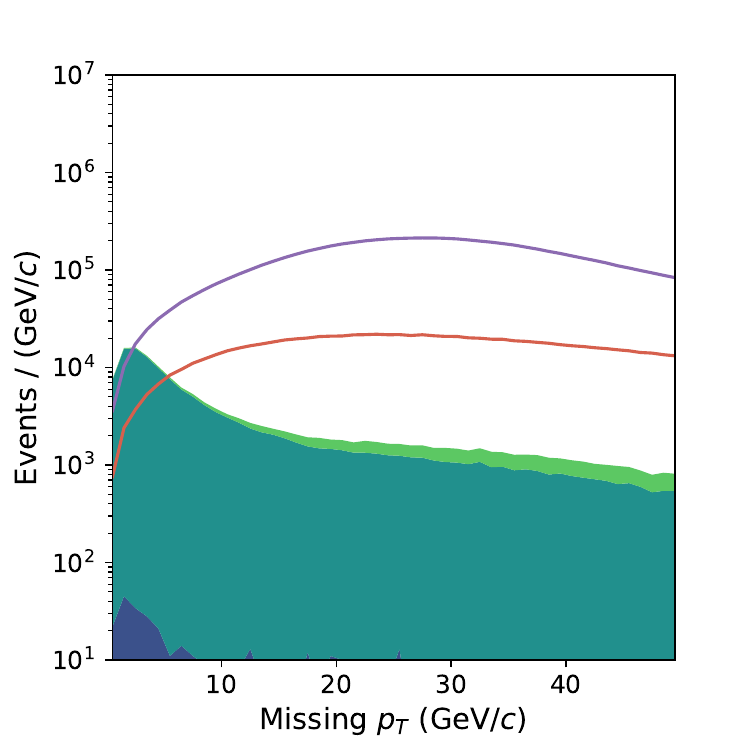}
    \includegraphics[width=7cm]{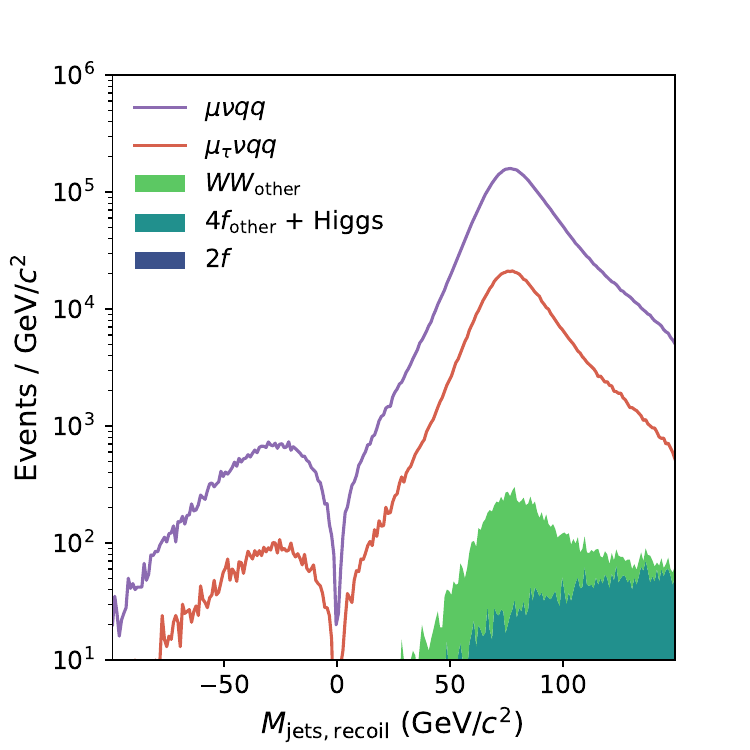}
    \caption{The distribution of variables in the cut flow of the muon channel. Similar to Fig.~\ref{fig:cutvariables}, the samples shown are the ones right before the relevant step in the cut flow.    
    }
    \label{fig:extracut1}
\end{figure}

\begin{figure}[htbp]
    \centering
    \includegraphics[width=7cm]{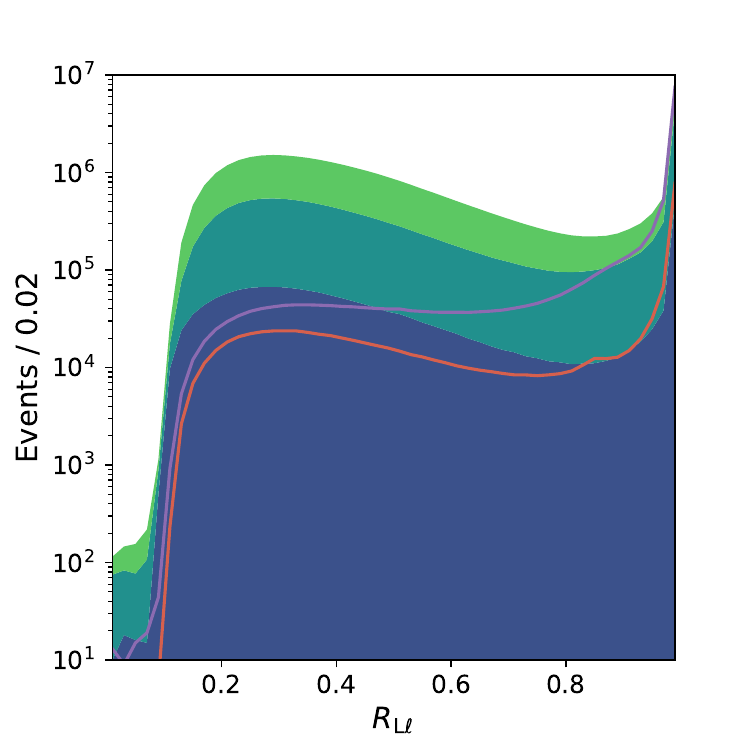}
    \includegraphics[width=7cm]{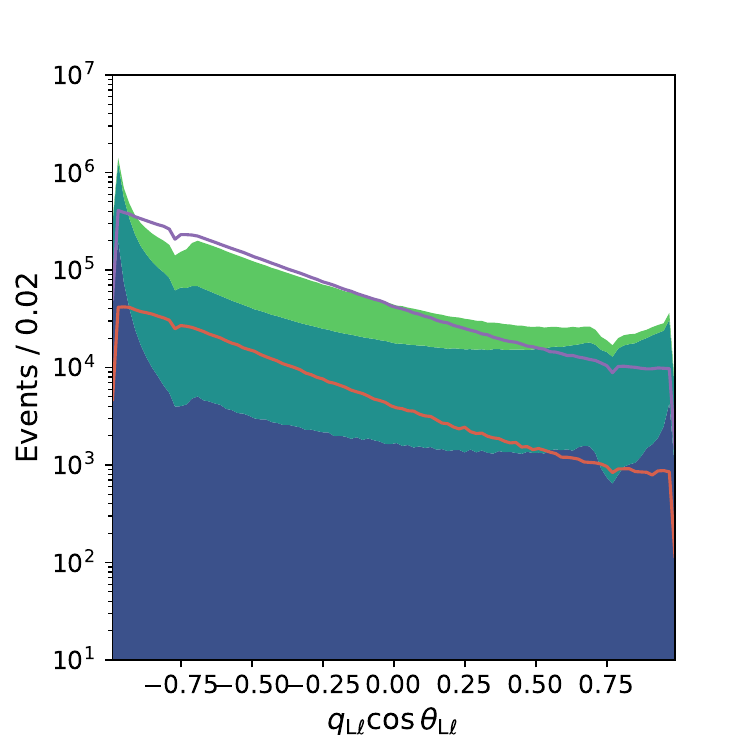}
    \includegraphics[width=7cm]{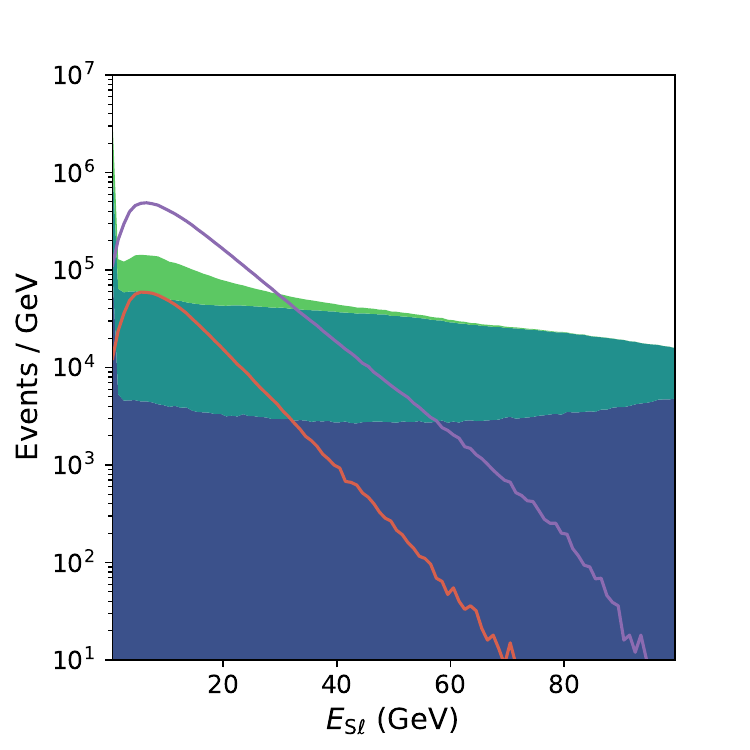}
    \includegraphics[width=7cm]{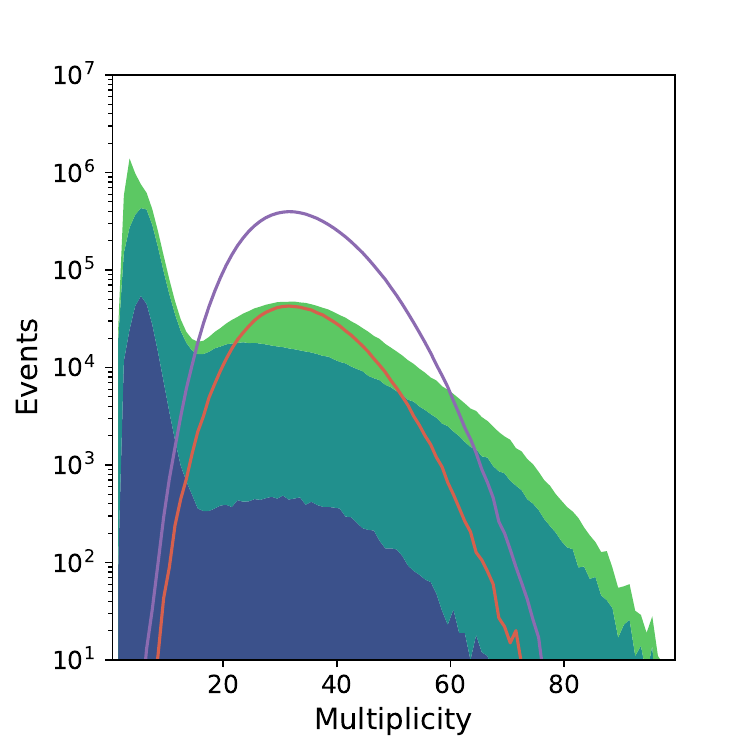}
    \includegraphics[width=7cm]{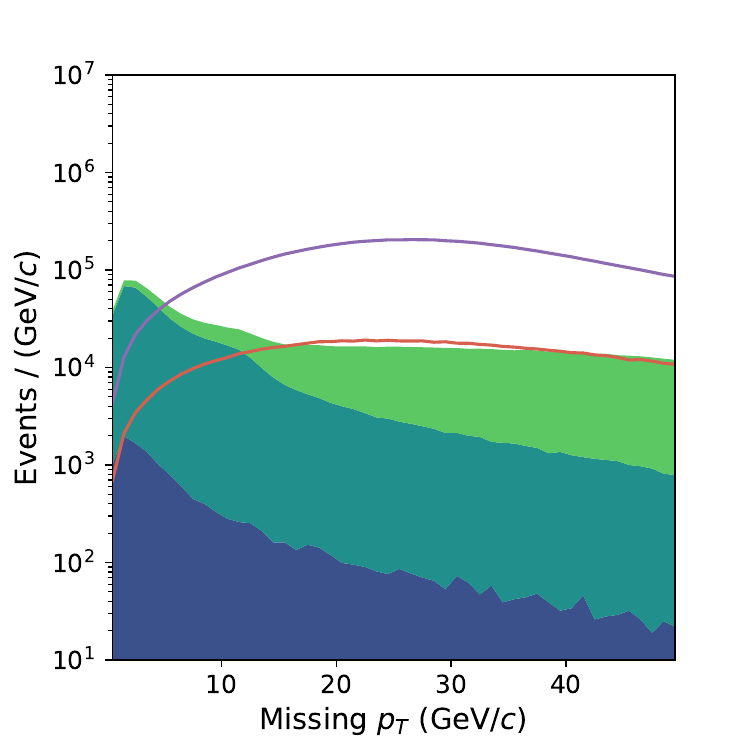}
    \includegraphics[width=7cm]{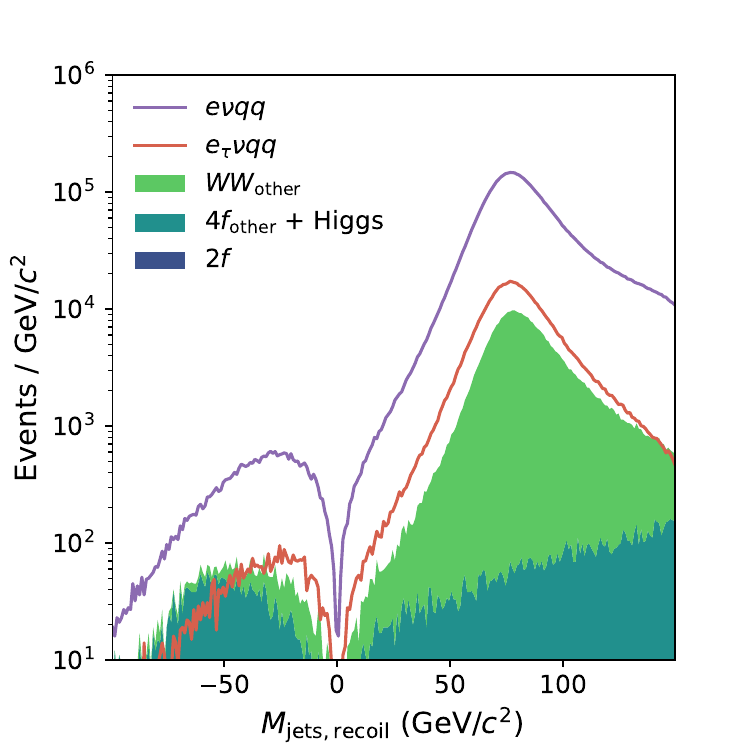}
    \caption{The distribution of variables in the cut flow of the electron channel. Similar to Fig.~\ref{fig:cutvariables_ele}, the samples shown are the ones right before the relevant step in the cut flow.}
    \label{fig:extracut2}
\end{figure}

Here we provide some of the featured kinematic distributions in the cut flow in Table~\ref{tab:evtselct1} and \ref{tab:evtselct2}. The muon and electron channel distributions are listed in Fig.~\ref{fig:extracut1} and~\ref{fig:extracut2}, respectively.

\section{Events Statistics Details}
\label{app:eventsstatistics}
Here we provide further details of the samples generated, especially their efficiencies in the cut flow. We start with the cut flow of the four sub-categories associated with the $WW_{\rm others}$ in the $e$ channel in Table~\ref{tab:WWother1}. Similarily, we also list the $WW_{\rm other}$ for the $\mu$ signal mode in Table~\ref{tab:WWother2}. Note that the signal events for one mode will be the backgrounds for another.

\begin{table}[h!]
\begin{center}
\small
\begin{tabular}{c|c|c|c|c}
\hline
$WW_{\rm others}$ for the $e$ mode &  $\mu\nu_{\mu} q\bar{q}$ &	$\mu_\tau \nu q\bar{q}$ &	$\tau(\rightarrow {\rm had.}+\nu)\nu q\bar{q}$ &	full had./lep. \\
\hline
w/o selections &48.5{{\tiny M}}&8.41{\tiny M}&31.4{{\tiny M}}&102.0{{\tiny M}}\\
$E_{{\rm L}\ell} > 12~{\rm GeV}$ &22.2{{\tiny M}}&3.83{\tiny M}&19.5{{\tiny M}}&70.4{{\tiny M}}\\
$R_{{\rm L}\ell} > 0.85$ &559{{\tiny K}}&63.2{{\tiny K}}&3.25{\tiny M}&17.6{{\tiny M}}\\
$\cos(\theta_{{\rm L}\ell})$ &559{{\tiny K}}&63.2{{\tiny K}}&3.25{\tiny M}&17.6{{\tiny M}}\\
$q_{{\rm L}\ell}\cos(\theta_{{\rm L}\ell}) < 0.00$ &414{{\tiny K}}&42.3{{\tiny K}}&2.78{\tiny M}&15.5{{\tiny M}}\\
2nd isolation $\ell$ veto &413{{\tiny K}}&42.3{{\tiny K}}&2.77{\tiny M}&14.6{{\tiny M}}\\
Multiplicity $ \ge 15$ &399{{\tiny K}}&40.1{{\tiny K}}&2.76{\tiny M}&528{{\tiny K}}\\
$\slashed{p}_T > 9.5~{\rm GeV}/c$ &387{{\tiny K}}&38.2{{\tiny K}}&2.64{\tiny M}&207{{\tiny K}}\\
$M_{\rm jets}>65~{\rm GeV}/c^2$ &364{{\tiny K}}&27.3{{\tiny K}}&2.57{\tiny M}&179{{\tiny K}}\\
$M_{\rm jets}<88~{\rm GeV}/c^2$ &188{{\tiny K}}&13.7{{\tiny K}}&1.46{\tiny M}&477\\
$M_{\rm jets, recoil} > -200~{\rm GeV}/c^2$ &188{{\tiny K}}&13.7{{\tiny K}}&1.46{\tiny M}&477\\
$M_{\rm jets, recoil}<115~{\rm GeV}/c^2$ &166{{\tiny K}}&7.7{\tiny K}&1.32{\tiny M}&184\\
$M_{{\rm L}\ell{\rm S}\ell} < 75~{\rm GeV} / c^2$ &163{{\tiny K}}&7.68{\tiny K}&1.29{\tiny M}&164\\
$\epsilon_{\rm kin} (\%)$ &0.3381&0.0912&4.1008&0.0002\\
\hline
    \end{tabular}
\caption{The kinematic cut flow for the four sub-categories of $WW_{\rm others}$ in the $e$ signal mode.}
\label{tab:WWother1}
\end{center}
\end{table}

\begin{table}[h!]
\begin{center}
\small
\begin{tabular}{c|c|c|c|c}
\hline
$WW_{\rm others}$ for the $\mu$ mode &  $e\nu_{\mu} q\bar{q}$ &	$e_\tau \nu q\bar{q}$ &	$\tau(\rightarrow {\rm had.}+\nu)\nu q\bar{q}$ &	full had./lep. \\
\hline
w/o selections &52.3{{\tiny M}}&8.66{\tiny M}&31.4{{\tiny M}}&102.0{{\tiny M}}\\
$E_{{\rm L}\ell} > 12~{\rm GeV}$ &791{{\tiny K}}&133{{\tiny K}}&687{{\tiny K}}&19.2{{\tiny M}}\\
$R_{{\rm L}\ell} > 0.85$ &7.01{\tiny K}&1.78{\tiny K}&71.7{{\tiny K}}&16.4{{\tiny M}}\\
$\cos(\theta_{{\rm L}\ell})$ &7.01{\tiny K}&1.78{\tiny K}&71.7{{\tiny K}}&16.4{{\tiny M}}\\
$q_{{\rm L}\ell}\cos(\theta_{{\rm L}\ell}) < 0.00$ &4.89{\tiny K}&1.32{\tiny K}&64.2{{\tiny K}}&14.3{{\tiny M}}\\
2nd isolation $\ell$ veto &4.89{\tiny K}&1.32{\tiny K}&64.2{{\tiny K}}&13.7{{\tiny M}}\\
Multiplicity $ \ge 15$ &4.59{\tiny K}&1.29{\tiny K}&63.9{{\tiny K}}&35.1{{\tiny K}}\\
$\slashed{p}_T > 9.5~{\rm GeV}/c$ &4.48{\tiny K}&1.22{\tiny K}&60.0{{\tiny K}}&25.0{{\tiny K}}\\
$M_{\rm jets}>65~{\rm GeV}/c^2$ &4.11{\tiny K}&955&58.2{{\tiny K}}&4.42{\tiny K}\\
$M_{\rm jets}<88~{\rm GeV}/c^2$ &711&618&40.7{{\tiny K}}&216\\
$M_{\rm jets, recoil} > -200~{\rm GeV}/c^2$ &711&618&40.7{{\tiny K}}&216\\
$M_{\rm jets, recoil}<115~{\rm GeV}/c^2$ &452&473&37.4{{\tiny K}}&37\\
$M_{{\rm L}\ell{\rm S}\ell} < 75~{\rm GeV} / c^2$ &452&473&37.4{{\tiny K}}&37\\
$\epsilon_{\rm kin} (\%)$ &0.0009&0.0055&0.1191&$<$ 0.0001\\
\hline
    \end{tabular}
\end{center}
\caption{Similar to Table~\ref{tab:WWother1} but for the $\mu$ signal mode.}
\label{tab:WWother2}
\end{table}

We also provide the detailed $\epsilon_{\rm kin}$ and $\epsilon_{\rm tag}$ for different signal and background processes in Table~\ref{tab:statistics_abs}. Note that the purity $\rho$ for each process is not well-defined, therefore we can only report the final event yield per ab$^{-1}$ instead of effective statistics.

\begin{table}[htb!]
    \centering
    \begin{footnotesize}
    \begin{tabular}{|c|ccccc|}
    \hline
      Analysis & Channel  & 
      Cross-section (ab) & $\epsilon_\text{kin}$ & 
$\epsilon_\text{tag}$ & 
Yields/$\text{ab}^{-1}$\\ 
    \hline
    \multirow{6}{*}{$\mu \nu cb$} & $\mu_{(\tau)} \nu cb$ & $2.4\times 10^3$ & 0.46 &  0.642 & 715	 \\
        & $\mu_{(\tau)} \nu cd/s$ & $1.4\times 10^6$ & 0.51 &  $1.1\times 10^{-4}$ & 91  \\
        & $\mu_{(\tau)} \nu qq_\text{other}$ & $1.4\times 10^6$ & 0.51 &  $1.5\times 10^{-5}$ & 11 	 \\
        & $WW_\text{other}$ & $9.7\times 10^6$ & $2.0\times 10^{-4}$ &  $6.1\times 10^{-4}$ & 1		 \\
        & $4f_\text{other}$ + Higgs & $6.9\times 10^6$ & $6.7\times 10^{-4}$
 &  $0.047$ & 20	 \\
        & $2f$ & $8.9\times 10^7$ & $1.9\times 10^{-6}$ &  0.029 & 4 \\
    \hline
    \multirow{6}{*}{$e \nu cb$} & $e_{(\tau)} \nu cb$ & $2.7\times 10^3$ & 0.37 &  0.43 & 449	 \\
        & $e_{(\tau)} \nu cd/s$ & $1.52\times 10^6$ & 0.39 &  $4.4\times 10^{-5}$ & 26 \\
        & $e_{(\tau)} \nu qq_\text{other}$& $1.52\times 10^6$ & 0.38 &  $5.0\times 10^{-6}$ & 3	 \\
        & $WW_\text{other}$ & $9.5\times 10^6$ & 0.0076 &  $3.9\times 10^{-4}$ & 28 \\
        & $4f_\text{other}$ + Higgs &$6.9\times 10^6$ & $2.0\times 10^{-4}$ &  0.0057 & 8 \\
        & $2f$ & $8.9\times 10^7$ & $6.4\times 10^{-5}$ &  0.0054 & 31 \\
    \hline
    \end{tabular}
    \end{footnotesize} 
    \caption{Statistics for each benchmark $WW\to \ell\nu qq$ mode, including contributions from $\tau\to \ell\nu\nu$ for signal.
    }
    
\label{tab:statistics_abs}  
\end{table}

\section{Alternative Tagging Methods and Determination of Efficiencies}
\label{app:traditionaltagging}
\begin{figure}[!htb]
    \centering
    \subfloat[]{ \includegraphics[scale=0.35]{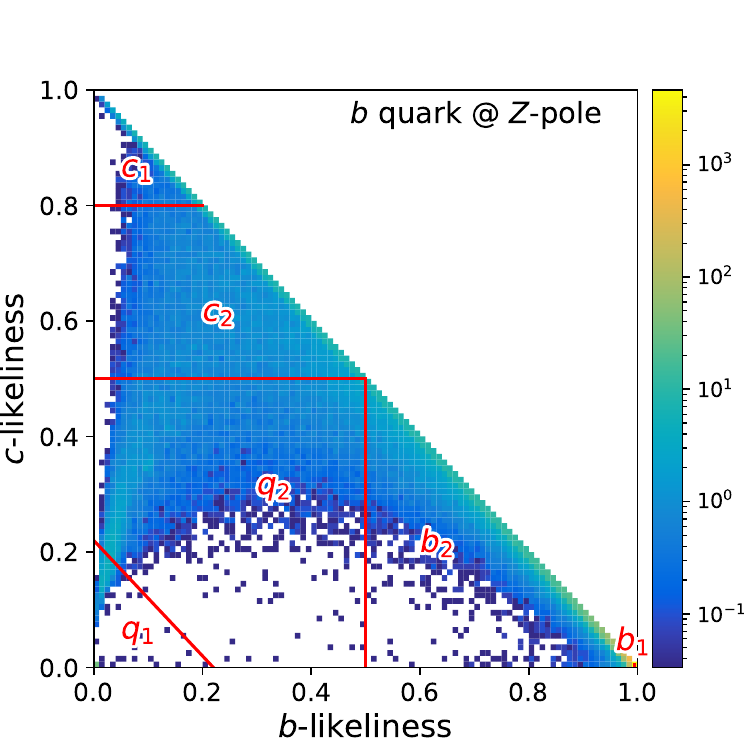}}
   \subfloat[]{  \includegraphics[scale=0.35]{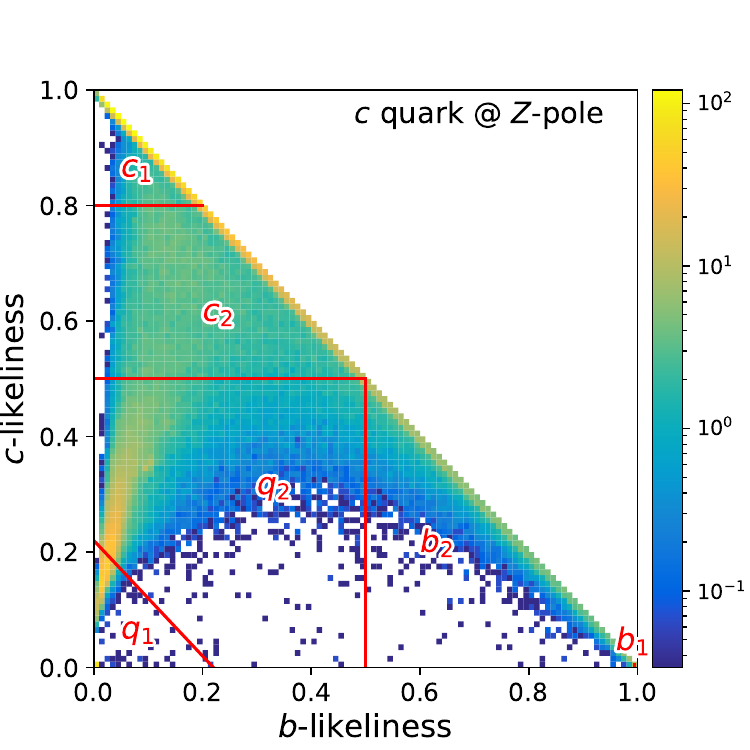} }
    \subfloat[]{ \includegraphics[scale=0.35]{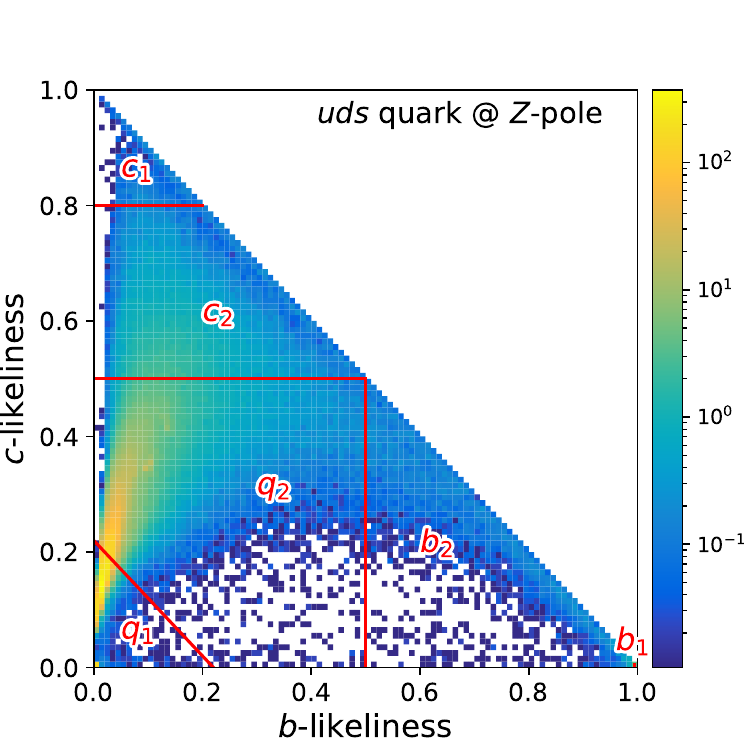}}
    \caption{flavor tagging performance at $Z$-pole and the definitions of labels for flavor tagging. The $b_1$ is corresponding to the region $b$-likeliness $>$ 0.995. }
    \label{fig:falvourtaggingperformance}
\end{figure}

Here, a more traditional tagging algorithm based on BDT is used, including multi-dimensional simulation inputs, e.g., track energy, displacement, and multiplicity. The output of this algorithm is two-dimensional, namely the $b$- and $c$-likeness, since the algorithm does not aim to identify lighter jets. As pointed out in~\cite{Faroughy:2022dyq}, having more working points of flavor tagging could substantially improve the sensitivity reach of tagging algorithms since they provide extra information by having more signal regions. In practice, each jet in the event is divided into six flavor-tagging categories, denoted as $b_1$, $b_2$, $c_1$, $c_2$, $q_1$, or $q_2$, based on the jet's $b$- and $c$-likeliness determined by the flavor tagging algorithm. The distribution of $b/c$-likeliness of different jets and the division of each flavor-tagging category is shown in Fig.~\ref{fig:falvourtaggingperformance}.

To reduce the accuracy requirement of those, we make a stringent cut on flavor tagging, i.e., $b$-likeliness $>$ 0.995. The algorithm's performance  using $Z$-pole dijet events for each category are listed in Table~ \ref{tab:flavortageff}.





\begin{table}[!htb]

    \centering
    \begin{tabular}{c|c|c|c|c|c|c}
    \hline
    quark $\backslash$ tag & $b_1$ & $b_2$ & $c_1$ & $c_2$ & $q_1$ & $q_2$ \\
    \hline
$b$ & 0.381&0.409&0.0175&0.109&0.0096&0.074\\
$c$ & 0.00049&0.083&0.228&0.354&0.070&0.264\\
$uds$ & 0.00025&0.0072&0.0013&0.059&0.430&0.502\\
    \hline
    \end{tabular}
    \caption{The flavor tagging efficiencies at $Z$-pole. The definitions of each tag labels are depicted in Fig.~\ref{fig:falvourtaggingperformance}.
    }
    \label{tab:flavortageff}
\end{table}



Applying the above algorithm to the samples after kinematic cuts with two jets generates total 21 possible combinations of labels. The statistical uncertainty are derived by the likelihood method, as there is no unique signal region defined. The overall likelihood $\mathcal{L}$ is defined as 
\begin{equation}
\label{eq:Likehood}
-2\log(\mathcal{L}) = \sum_{i=1}^{21} \frac{(\sum_a S_a N_i^a + N^{\rm bkg}_i - N^{\text{obs}}_i)^2}  {N^\text{obs}_{i} },
\end{equation}
where $N_{i}^a$ is the expected number of events for the event type $a$ as predicted by SM.
$S_a$ is the signal strength of signal $a$.
Signal types include $\mu_{(\tau)}\nu cb$, $\mu_{(\tau)}\nu ud/s$, and $\mu_{(\tau)}\nu cd/s$.
$N^\text{obs}_{i}$ is the number of observed event in category bin, $i$.
$N^{\rm bkg}_i$ is the number of background events in category bin, $i$.
The statistical uncertainty of $R_{cb}$ is estimated by the variation of $\mathcal{L}$ with various $W$ hadronic decay signal strengths. The stastical uncertainty of $|V_{cb}|$ from the muon channels at an unpolarized Higgs factory reaches 1.5\%(0.75\%) for the $L = 5(20)$~ab$^{-1}$ case.

\bibliographystyle{JHEP}
\bibliography{references}
\end{document}